\documentclass[USenglish,oneside,twocolumn]{article}

\usepackage[utf8]{inputenc}
\usepackage[big]{dgruyter_NEW}

\usepackage{amsmath,amssymb,amsfonts}
\usepackage{listings,listings-rust}
\usepackage{booktabs}
\usepackage{graphicx}

\usepackage{textcomp}
\usepackage{multicol}

\usepackage{float}
\usepackage{hyperref}
\usepackage{algorithm}
\usepackage{algpseudocode}

\usepackage{tcolorbox}
\usepackage{pifont}%
\usepackage[normalem]{ulem}

\usepackage{array}
\usepackage[justification=centering]{subcaption}
\usepackage{makecell}
\def\BibTeX{{\rm B\kern-.05em{\sc i\kern-.025em b}\kern-.08em
    T\kern-.1667em\lower.7ex\hbox{E}\kern-.125emX}}
    
\usepackage{xcolor}
\definecolor{codegreen}{rgb}{0,0.6,0}
\definecolor{codegray}{rgb}{0.5,0.5,0.5}
\definecolor{codepurple}{rgb}{0.58,0,0.82}
\definecolor{backcolour}{rgb}{0.95,0.95,0.92}

\lstdefinestyle{mystyle}{
    backgroundcolor=\color{backcolour},   
    commentstyle=\color{codegreen},
    keywordstyle=\color{magenta},
    numberstyle=\tiny\color{codegray},
    stringstyle=\color{codepurple},
    basicstyle=\ttfamily\scriptsize,
    breakatwhitespace=false,         
    breaklines=true,                 
    captionpos=c,                    
    keepspaces=true,                 
    numbers=left,                    
    numbersep=4pt,                  
    showspaces=false,                
    showstringspaces=false,
    showtabs=false,                  
    tabsize=2
}
\newcolumntype{L}[1]{>{\raggedright\arraybackslash}p{#1}}
    
\cclogo{\includegraphics{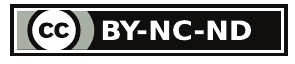}}
\begin{document}

\newcommand{\michal}[1]{{\em\color{purple}{[#1 --- michal]}}}
\newcommand{\fix}[1]{{{#1}}}
\newcommand{\fixfix}[1]{{{#1}}}
\newcommand{\lucek}[1]{{\color{green} \textbf{Lucek:} #1}}

\newcommand{\dali}[1]{{\em\color{red}{[#1 ---dali]}}}
\newcommand{\ian}[1]{{\em\color{orange}{[#1 ---ian]}}}

\newcommand{\todo}[1]{}
\newcommand{\corrauth}{Corresponding Author}

\newcommand{\cmark}{\ding{51}}%
\newcommand{\xmark}{\ding{55}}%
\newcommand{\khandle}{{\sf key\_handle}}
\newcommand{\khandles}{{\sf key\_handles}}
\newcommand{\uid}{{\sf uID}}
\newcommand{\mA}{\mathcal{A}}
\newcommand{\mB}{\mathcal{B}}
\newcommand{\mC}{\mathcal{C}}
\newcommand{\mR}{\mathcal{R}}
\newcommand{\exec}{\leftarrow}
\newcommand{\getAssert}{{\sf authenticatorGetAssertion }}
\newcommand{\allCred}{\textit{allowCredential }}
\newcolumntype{x}[1]{>{\centering\arraybackslash\hspace{0pt}}p{#1}}

  \author*[1]{Michal Kepkowski}

  \author[2]{Lucjan Hanzlik}

  \author[3]{Ian Wood}

  \author[4]{Mohamed Ali Kaafar}

   \affil[1]{Macquarie University, E-mail: michal.kepkowski@students.mq.edu.au}

  \affil[2]{CISPA Helmholtz Center for
Information Security, E-mail: hanzlik@cispa.de}

  \affil[3]{Macquarie University, E-mail: ian.wood@mq.edu.au}

  \affil[4]{Macquarie University, E-mail: dali.kaafar@mq.edu.au}

\title{How Not to Handle Keys:
Timing Attacks on FIDO Authenticator Privacy}


\begin{abstract}
{
This paper presents a timing attack on the FIDO2 (Fast IDentity Online) authentication protocol that allows attackers to link user accounts stored in vulnerable authenticators, a serious privacy concern.
FIDO2 is a new standard specified by the FIDO industry alliance for secure token online authentication. It complements the W3C WebAuthn specification by providing means to use a USB token or other authenticator \fix{(which holds the secret authenticating material and implements FIDO protocols)} as a second factor during the authentication process. From a cryptographic perspective, the protocol is a simple challenge-response where the elliptic curve digital signature algorithm is used to sign challenges. To protect the privacy of the user the token uses unique key pairs per service. 
To accommodate for small memory, tokens use various techniques that make use of a special parameter called a \emph{key handle} sent by the service to the token  
\fix{with which the token can securely produce an authentication key (through generation or decryption). }
We identify and analyse a vulnerability in the way 
the processing of key handles is implemented that allows attackers to remotely link user accounts on multiple services. 
We show that for vulnerable authenticators there is a difference between the time it takes to process a key handle for a different service but correct authenticator, and for a different authenticator but correct service. This difference can be used to perform a timing attack allowing an adversary to link
\fix{user's accounts across services. We present several real world examples of adversaries that are in a position to execute our attack and can benefit from linking accounts. We found that}
two of the eight hardware authenticators we tested were vulnerable despite FIDO level 1 certification, indicating a not insignificant problem. 
This vulnerability cannot be easily mitigated on authenticators because, for security reasons, they usually do not allow firmware updates. 
In addition, we show that due to the way existing browsers implement the WebAuthn standard, the attack can be executed remotely. 
However, we discuss countermeasures that can be implemented by 
browser providers to mitigate the remote form of the attack.
}
\end{abstract}

\maketitle

\section{Introduction}
\todo{7. [B,C,D] The reviews note a number of shortcomings in the paper's grammar and copyediting that need to be addressed in the revision submitted.
}
\todo{Add in intro the importance of L3 certification}



\noindent

Password-based authentication methods are infamous for their security weaknesses \cite{DBLP:conf/ccs/ThomasLZBRIMCEM17,DBLP:conf/soups/KarunakaranTBC18}, which led to the adoption of second factor authentication such as software based approaches like Google Authenticator (https://g.co/2step) and Duo Security (https://duo.com), and hardware based tokens like Yubico Yubikey (https://www.yubico.com/) and HyperFIDO (https://www.hypersecu.com/).   
Authentication tokens provide a challenge-response based protocol using a standard specified by the FIDO Alliance \cite{FIDO2} called FIDO2 (Fast Identity Online), as a successor of UAF (Universal Authentication Framework) \cite{UAF} and U2F (Universal 2nd Factor) \cite{U2F}. 
%
%
%
The authenticator/token holds a secret key that is used to authenticate against a public key bound to the user's account during registration. FIDO2 with its  Client to Authenticator Protocol (CTAP)~\cite{FIDO_ctap_spec}, combines both and is the state-of-the-art standard for user-side authentication which is complemented by the W3C WebAuthn specification \cite{Kumar:21:WAA}.

The adoption of FIDO2 is driven by the support of major service providers on both mobile and desktop platforms (Android, IOS and Windows \cite{FIDO_android_announcement}\cite{FIDO_windows_announcement}\cite{FIDO_ios_announcement}) and industry implementations (some examples are: US login.gov page \cite{FIDO_deployment_us_gov}, ID intelligence suite from VISA \cite{FIDO_deployment_visa}, NHS enhanced login \cite{FIDO_deployment_nhs}).
%
%
The main goal of the FIDO2 protocol is to mitigate known problems with existing authentication mechanisms. In particular, the challenge-response design of the protocol protects users against replay and credential theft attacks. The protocol is known to be immune to database leaks since the authentication servers store only the public keys of users \cite{FIDO_webauthn_spec}. 
The protocol is also designed to protect the privacy of users by preventing the linkage of accounts for which the same authenticator was registered (see "Privacy considerations for authenticators" section of WebAuthn \cite{FIDO_webauthn_spec}).
While most of the research efforts into FIDO and related technologies have focused on providing security guarantees, in particular with strong and robust end-user authentication, considerations of privacy have received less attention from the security community.

The linking of a users' accounts across internet platforms, our focus here, poses both common risks such as undesirably targeted recommendations and advertising~\cite{4cefd0d46505406a96e10ae7d7afcaae} and other purposes \cite{DBLP:conf/www/GogaLPFST13,DBLP:conf/sp/NarayananPGBSSS12,DBLP:conf/www/RiedererKCKL16}, \fix{as well as} more dangerous risks such as enabling actors with malicious intent towards an individual such as criminals or \fix{unfriendly} state actors.
\fix{FIDO2 authenticators are one of the proposed solutions to these problems and are being actively promoted by security oriented companies. For example, Microsoft, Yubico and Google run programs to empower at-risk individuals by ensuring strong and privacy preserving authentication via FIDO2, and in 2021, they distributed 35k tokens~\cite{noauthor_yubico_2021,noauthor_google_2021}.
}

In a regular login/password based authentication scenario, a user can aim at protecting their privacy across services by creating a service-specific identity, \fix{e.g.,~}by using different e-mail accounts during registration and different login/password information. 
%
Even though the method is not perfect, it allows some protection against providers trying to easily link user accounts across different services. Note that a malicious service can always use other metadata \fixfix{(e.g., tracking cookies or geolocation)} to link user accounts and protection against such attacks is an orthogonal problem.  
Unfortunately, the introduction of \fixfix{any} second factor for authentication \fix{increases} the privacy threat of linking identities across services. In particular, binding the same hardware token with unique public keys to the user's accounts connects the identity of the  user across the various services. \fixfix{The FIDO2 protocol seeks to mitigate this risk through privacy preserving registration and authentication algorithms.}

Unlinkability across services together with user authentication is one of the two  \emph{fundamental} properties of the FIDO2 authentication mechanism~\cite{fido-security}. In fact, the FIDO standard recommends that hardware tokens generate unique public keys for each service. Unfortunately, in practice this solution does not scale well if the device has only a limited amount of secure memory to store secret keys. 
This issue is addressed by the consideration of \emph{non-resident} keys, \fix{i.e.:~}keys for which the secret key is recomputed during the authentication process and not stored on-device. A common approach to implement this is to use the \emph{key-wrapping} technique \cite{9176}. The idea is for the server to provide the token with a ciphertext containing the secret key that can be used for authentication. This secret key is then decrypted using a master key stored on the device.
We show that wrong processing of key handles can lead to a significant difference in execution time depending
on what data is used. 
In particular, \fix{ for vulnerable authenticators there is a time difference between processing key handles from a given authenticator and those that are not. If the attack is performed in the context of a valid user authentication, this provides the attacker with an approach to link a candidate key handle (and its associated user account) with the key handle and account to which the user is authenticating. } 
Note that Relying Parties, the services to which a user authenticates, may stipulate that a \emph{resident key} is required, which quickly consumes the memory of authenticator tokens and thus reduces their utility, however renders the site \fix{immune} to this attack. 

\fix{We focus on a remote form of the attack in which malicious software running on the users hardware is not required, but the attacker must have the capability to modify FIDO communications and to time FIDO calls to the authenticator. In practice, the timing must  be done by malicious JavaScript code, so either ownership of JavaScript related to authentication or the ability to inject JavaScript is required. }
\fix{Surprisingly, numerous parties that interact with FIDO flow have these capabilities. We present a list of such actors with corresponding motivations and attack scenarios, including FIDO service providers, web proxies and adversaries exploiting XSS (Cross Site Scripting) vulnerabilities.
Note that malicious code running on user hardware with a CTAP API (e.g.,~malicious apps or compromised browsers on Linux, Windows and MacOS, a rather stringent requirement) can execute silent CTAP authentications to both probe authenticators to link key handles and actually authenticate with user services when key handles are known. }

The FIDO2 specification stipulates that an authenticator must perform a user presence check during the first phase of authentication, adding an indeterminate delay to the corresponding CTAP call. 
We propose two approaches to mitigate this and thus allow the timing attack to proceed. 
Our first proposal utilises multiple key handles in a single CTAP call that requires only a single user presence check. In this way, the indeterminate nature of the user response can be averaged over multiple calls rendering the timing difference measurable. 
This attack is close to undetectable but requires clients (typically browsers) to allow long lists of key handles in a single CTAP call, which is the case for several major web browsers. 
%
Our second proposal uses audio recording to identify the point in time at which a physical button on the authenticator is pressed, thus circumventing restrictions in some clients on the number of handles in a single CTAP call. We demonstrate that the button click is easily detectable in a typical home environment. 

\todo{[B,D] The reviewers were hoping for more clarification of the novelty of the techniques. While it is ok if the individual techniques themselves are not fully novel, the reviewers were hoping for the authors to provide an appropriate and measured sense of what they feel the important combination of techniques or new insights thereof are in this work. I want to emphasize that no overclaiming is needed, but instead to encourage the authors to think more clearly about any generalizable lessons they hope the readers take away.}


%
%
\fix{We perform experiments} with popular FIDO2 clients running on different operating systems and authenticators from various manufacturers. 
Our analysis reveals that two of the eight hardware authenticators we tested are vulnerable to our timing attacks. Moreover, our experiments show that \fix{the attacks} can be executed remotely (for example as an external web service) through popular web browsers (Chrome, Firefox, Safari and others). 
We note that the FIDO security measures document stipulates in [SM-29]: ``No leakage of secret information to remote entities via variation of operation execution time'' \cite{fido-security}. The vulnerable tokens, which have both passed a security certification on L1 by the FIDO Alliance, clearly fail in this respect. \fix{Our examination of the certification procedures revealed that only L3 certification provides in-depth and on-device testing \fixfix{(i.e., empirical tests executed on the authenticator)}, however at the time of writing there are no authenticators with L3 certification.}

One might hope that the proposed attacks can be easily mitigated by providing a firmware update for vulnerable authenticators. Unfortunately, to increase security most come without an update functionality. As a partial solution that prevents remote attacks, we propose and discuss ways for browser providers to mitigate the attacks. Note that the attack would still be possible from software running on the users device, hence replacing vulnerable tokens is the only complete solution. 

To better understand the extent of the threat, we developed and ran a web crawler, which gathered information about FIDO2 implementations in the wild. We collected 684 records of FIDO2 in Javascript sources from high traffic websites. 
The results showed that the only the passwordless implementation used by Microsoft \fix{required resident keys} and could not be used for deploying the attack. The remaining implementations use FIDO2 as a second factor with non-resident keys and are thus in position to break the privacy of vulnerable FIDO authenticators. 

\fix{Our approach leverages previously undiscovered weaknesses in the FIDO2 specifications and the ways that \fix{those specifications} are implemented. 
Drawing on well-established techniques to exploit side-channel vulnerabilities such as improper error handling and execution time differences, we succeeded in finding a \fixfix{novel and} easy to execute chain of actions that allow an adversary to learn additional information from vulnerable authenticators. We discuss the consequences of our findings as well as lessons applicable to any authentication system.}

The contributions of this paper can be summarized as follows:
\begin{enumerate}
    \item We present a timing attack on the FIDO2 protocol that enables attackers to link user accounts, a serious privacy breach. 
    \item We demonstrate a remote execution method that allows the attack to be performed by website JavaScript code. 
    \item Two of the eight hardware token authenticators we tested were vulnerable out of a field of 111, indicating a substantial public privacy concern. 
    \item We proposed mitigation measures for FIDO clients that prevent the remote form of the attack and for FIDO authenticators. 
    We notified relevant vendors and we participated in the mitigation design.
    \item We surveyed 1 million high traffic web sites and found 684 FIDO authentication deployments, of which almost all allow \emph{non-resident} keys and are thus exposed to our attack.
\end{enumerate}

\subsection{Responsible Disclosure}
\todo{8. [Reviewer discussion] Before publication, the reviewers want the authors to verify that responsible disclosure of this vulnerability was followed.}
In accordance with ethical standards we notified the relevant FIDO authenticator and client vendors about our findings prior to submission at \fix{least} 10 weeks prior to submission and informed them of the expected minimum date of publication. 
We sent our report to Chromium (which implements \fix{the} core engine for Chrome, Edge and Opera browsers), Firefox, and Safari teams. We informed Hypersecu \fix{and} Feitian and requested that they contact us should there be any concerns. We also reached out to the FIDO Alliance with details of our findings. 
Details of our communications and their responses can be found in Appendix \ref{Apdix:vendors}. 

\section{FIDO Authentication}
\subsection{High Level Overview}

The process is executed 
between 3 parties: an authenticator (\fix{i.e.,~}the hardware token), 
a client (\fix{i.e.,~}a browser) and a FIDO server. 
From a cryptographic perspective, it is a simple challenge-response protocol where the server issues a challenge to which the authenticator responds in form of a digital signature (also called \emph{assertion}) on the challenge and other session data including the server's origin provided by the client. 

The FIDO server is a backend element of the service which performs validations of assertions. It might be an incorporated element of the Relying Party (RP) or it can be an external service. The user's platform is running a client that acts as a proxy between the authenticator and the server. The communication channel is constructed using two specifications: CTAP \cite{FIDO_ctap_spec} and WebAuthn \cite{FIDO_webauthn_spec}. The former defines an API to exchange messages with the authenticator using a low-level protocol with CBOR encoded messages. The CTAP specification allows the usage of USB, NFC, and BLE technologies for the transportation layer. WebAuthn defines an API that specifies how web-applications should call client-specific functions. In particular, the API defines a javascript-specific navigator object that can be used to issue a request for registration (called credential creation) or authentication to the client which in turn uses CTAP to send the request to the authenticator.
\begin{figure}[ht]
\centering
    \includegraphics[width=0.9\columnwidth]{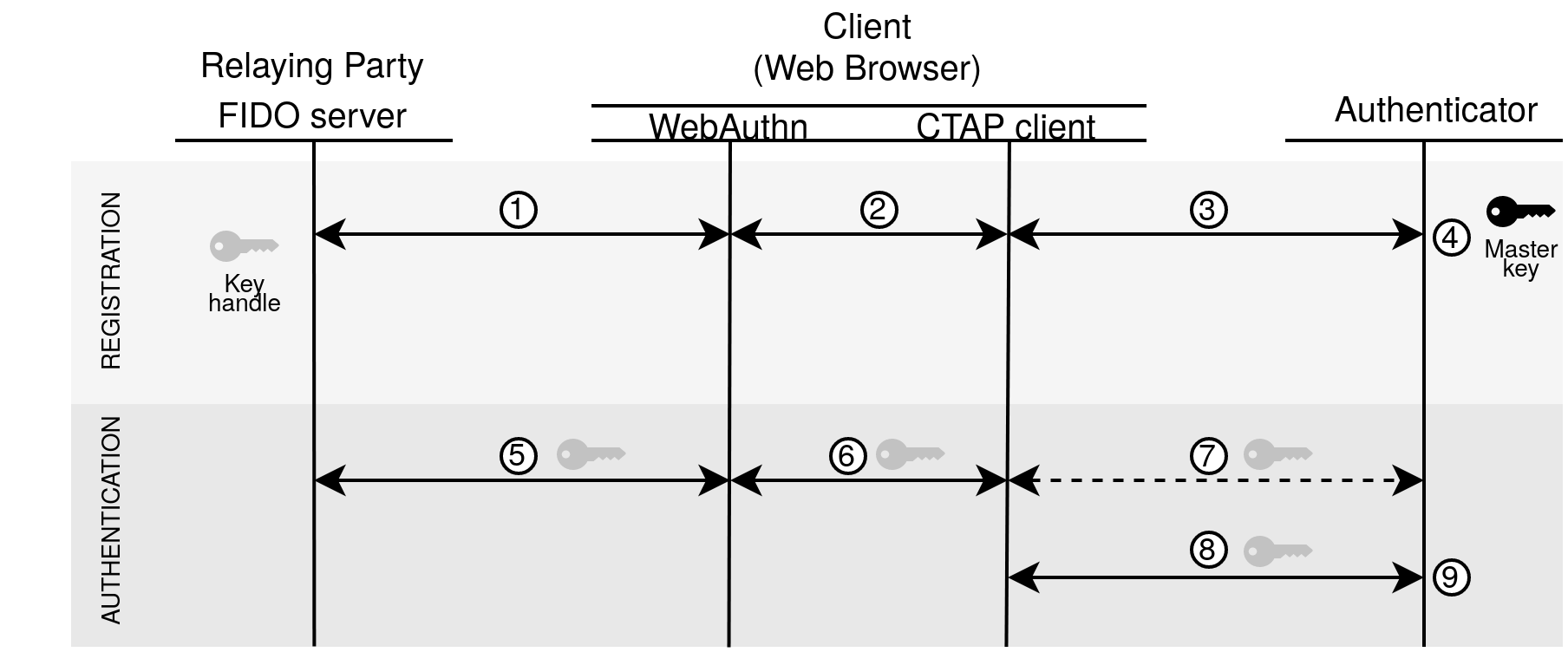}
    \caption[Short Title]%
    {FIDO2 simplified flow of registration and authentication. }
    \label{fig:flow}
\end{figure}
A simplified flow of the protocol is given in \fix{F}igure~\ref{fig:flow}, where:
\begin{enumerate}
    \item RP calls the \textit{navigator.credentials.create} method.
    \item Browser translates Webauthn call to CTAP request.
    \item CTAP client calls \textit{authenticatorMakeCredential}.
    \item Authenticator generates key pair and key handle, the response is returned back to RP.
    \item RP initiates by calling \textit{navigator.credentials.get} with key handle.
    \item Browser translates Webauthn call to CTAP request. 
    \item Silent authentications to check key handles.
    \item CTAP client calls \textit{authenticatorGetAssertion}.
    \item Authenticator verifies key handle, performs required user checks and signs the assertion which is returned back to RP.
\end{enumerate}

\subsection{FIDO Authentication - Step by Step} \label{sec:stepByStep}

In the FIDO specification we distinguish two phases: \emph{registration} and \emph{authentication}. We will now describe this process for a typical use case where the client is a browser, the authenticator is a USB hardware token (for example implementing ECDSA --- Elliptic Curve Digital Signature \cite{journals/ijisec/JohnsonMV01}) 
and the relying party is a standard web-server.
While we overview the entire process, we focus on the details that are relevant to the execution of the timing attack. 

\subsubsection{Registration}
The purpose of this phase is to bind the authenticator to the user's account. Similar to the standard login/password based scenario the user is provided an interface to send out registration requests to the server. 
Prior to receiving such a request the server generates a unique challenge value and returns it to the browser which handles the whole registration process using a server-provided  JavaScript based application. The browser then uses credential management API which internally executes the CTAP protocol with the authenticator. 
The challenge and additional data like the server origin in form of an application ID are sent to the authenticator using (in our scenario) USB.



After the user presence check, the authenticator internally generates a key pair for the ECDSA signature scheme and uses the secret key to create the server's challenge. The client's request specifies whether the secret key should reside on-device or the key pair is of the non-resident type mentioned before. In the latter case, additionally to the response, the authenticator also returns something called a key handle. Depending on the implementation of the non-residents key this can be either a random value that is used as input to a KDF (key derivation function) or the encryption of the secret key for ECDSA. 

Finally, the public key, key handle, assertion, and optional attestation (of the public key) are sent to the server. The data is then verified and if accepted the public key and key handle are added to the database.

\subsubsection{Authentication}
After successful registration, the user now tries to access her account and is prompted to enter login information. This first step allows the server to locate the user's account and the devices bound to it. 
Note that it is commonly allowed to register multiple devices with a single server. A list (also called allowed list) of key handles corresponding to those devices is sent to the client and used as input
to the \verb+navigator.credential.get(...)+ API call. Since the list 
provided by the server contains key handles for allowed devices the browser tries to find the right key handle. This is internally done by asking the hardware token to authenticate giving each element on the list as input. Note that if those key handles correspond to different devices there will be only one key handle that will generate a valid assertion. Interestingly enough this internal verification is performed \emph{without} user presence check since otherwise the user would be required to click a button a couple of times (depending on the size of the allowed list). After the right key handle is found the browser issues the last call to the token and requires a user presence check.
%
Finally, the authenticator creates the assertion which is a standard ECDSA signature on a message containing among others the server's challenge, an authentication counter (to protect against cloned devices), and the origin. The browser sends the assertion to the server that provided access to the service if the verification was successful.

\subsection{FIDO2 Security}
The design of \fix{the} FIDO2 protocol aimed to mitigate the known vulnerabilities of other authentication mechanisms. The FIDO2 core functionality is implemented using public key cryptography which
enables the following key features. Firstly, the RP's database holds only public keys which makes FIDO2 resistant against database leakage. Additionally, the protocol requires the FIDO server to use a random and unique challenge for each transaction which protects against 
replay attacks.

Another important security feature provided by FIDO2 is phishing resistance. Over the years phishing has grown in popularity as an easy to execute malicious activity \fix{and has been shown to be} 
the second largest group of malicious activities~\cite{abs-1904-10629}. 
Combined with  MITM attacks, it is a major threat, especially in working from home environments. 
Such attacks are not mitigated by popular second factor mechanisms such as SMS or OTP codes. FIDO2 addresses this threat by including the relying party origin parameter into the authentication process. 

The FIDO Alliance places an emphasis on ensuring the privacy of users is preserved \cite{FIDO_privacy_principles,webauth_priv_considerations}. 
In FIDO2 this is possible because \fix{for each registration a new credential is generated that is not linked to the user.} 
\fix{There is no privacy attack on FIDO2 protocol that we know of.}

\fix{
\subsection{FIDO Alliance certification}
Verification of security and privacy guarantees of FIDO2 parties is a challenging and demanding process. \fix{The} FIDO Alliance introduced an unified and voluntary program to certify products against a list of security and privacy controls. In particular, a dedicated certification path was designed for FIDO2 authenticators. At the time of writing, there are three main levels (L1, L2 and L3) and recently added "plus" levels (L1+ and L3+). Levels define a set of required controls and testing methodologies \cite{FIDO_cert_requirements}. Security and privacy are checked on all levels, however, advanced on-device testing techniques such as testing for timing vulnerabilities are only executed for L3 certification. Verification actors differ between levels: For L1, evaluations are completed by the FIDO Security Secretariat, whereas for L2 and L3 they are executed by FIDO Accredited Security Laboratories. Currently, the majority of the certified authenticators reached only L1 (134). L2 is achieved by 9 authenticators and there are no devices with L3.}

    

\section{Adversarial Model and Attack} \label{sec:concept}
\todo{1. [A,B,C,D] A major critique of this paper was that the adversarial model was often not clearly described, and perhaps described partially incorrectly. The reviewers especially wondered whether a stronger model than the authors described was actually necessary (see Review A). The reviewers expect the revision will substantially restructure section 3 to make the adversarial model more clear, as well as describe the adversarial model accurately relative to the potential issues raised by the reviewers.}


\fix{
FIDO2 authentication scenarios require a certain level of flexibility (e.g., allowing users to register multiple authenticators under the same account). The FIDO2 protocol provides mechanisms for additional functionalities to support these scenarios. Two of them, which are particularly useful for our attack, are silent authentication and the \allCred list. Implementation flaws of these functionalities, described below, lead to a remote channel to measure authenticator execution time. This capability enables an adversary to detect differences in key handle processing times. Notably, the most popular mechanism to store key handles (non-resident keys), if incorrectly implemented, can allow a timing side channel to identify key handles that are associated with a given authenticator. Given that authenticators are typically used by single individuals, the combination of the elements mentioned above creates a remote attack vector that allows an adversary to achieve the goal of linking an individuals FIDO registrations. 

In sections below, we provide detailed descriptions of the features that lead to the attack, then present an adversarial model and attack algorithm. Finally, we provide several examples of possible adversaries. }

\subsection{Remote CTAP Calls and Webauthn API Implementation}

To describe how an adversary can \fix{measure execution time differences} without user interaction we first have to explain two different types of user checks defined by the FIDO2 specification. The first one is \emph{user presence} which requires the authenticator to check if a human actor is present to proceed with authentication. It is worth noting that popular implementations known from hardware tokens can be easily simulated (e.g., button click can be done \fix{by} machine). The second check is called \emph{user verification} and it aims to authorize the accepting party. There exists a variety of available implementations such as memorized secrets (e.g., PIN code) or biometric authentication (e.g. fingerprint). 

User presence and user verification are configurable in the CTAP protocol as simple flags. User presence is always required during registration but the FIDO2 specification allows the CTAP client to modify both flags during authentication/assertion. \fix{This} means that the CTAP client can trigger assertion\fix{s} without user input (also called silent authentication). \fix{Silent} authentication does not usually trigger any indication on the tokens themselves. \fix{E.g.,} the LED indicator \fix{that usually} signals that the token is waiting for human action 
remains unchanged. It is easy to imagine a scenario where malicious software is probing a hardware token left connected to the users platform.

Silent authentication is a \fix{useful} mechanism for FIDO2 clients, since they can use it to filter key handles in the \allCred list provided by \fix{the} WebAuthn API. The user is not bothered to provide a user presence check for each attempt and a CTAP client \fix{(e.g.,~a browser)} can identify which key handle belongs to the \fix{authenticator}. After finding the correct \fix{key handle} the CTAP client continues with the assertion that requires user presence or verification. 
This functionality can be abused by sending an \allCred list with key handles to check. In other words, an adversary can remotely \fix{trigger the} execution of silent authentications on the token by creating a proper combination of key handles in the \allCred parameter provided by the WebAuthn API. 



\begin{figure}[!ht]
    \centering
    \includegraphics[width=\columnwidth]{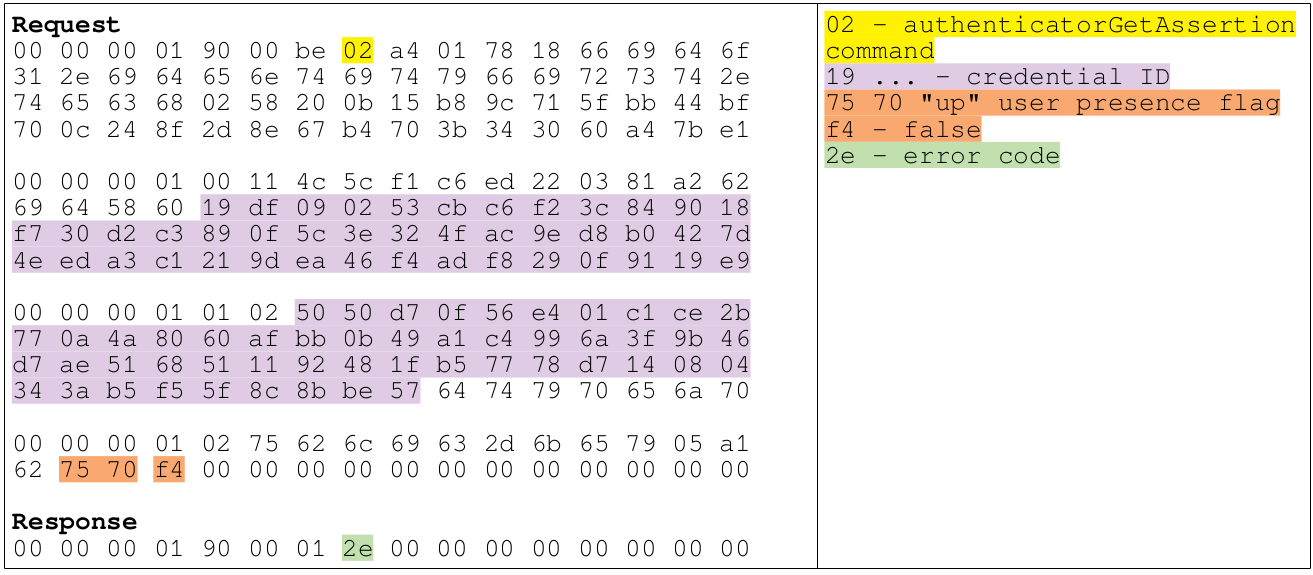}
    \caption{Data exchanged in a single CTAP silent authentication captured by intercepting USB traffic using Wireshark software triggered from Chrome browser}
    \label{fig:ctap_data}
\end{figure}

\fix{We found that all browsers considered in this study (Chrome, Brave, Firefox, Opera, Edge and Safari) parse an \allCred parameter with multiple key handles in this way. Analysis of the Chromium browser source code (used in many commercial browsers, e.g.,~Chrome, Opera and Edge) revealed that silent authentication is executed whenever the \allCred list contains more than one element (see \cite{chromium-fido} line 224-232). 
We further examined the data exchanged with a USB hardware token for the Firefox and Chrome browsers, which clearly showed silent authentications for each key handle in the \allCred list are made until a key handle successfully authenticates, after which a non-silent authentication with the found key handle was triggered. 
}\fixfix{The remaining browsers (Brave, Edge and Safari) perform a single user presence check when there are multiple entries in the \allCred list, indicating that silent authentications are also used in this way. }
In Figure~\ref{fig:ctap_data} we present data exchanged for a single silent authentication and highlighted the most important parts of the message.

The last obstacle the adversary needs to overcome is a user presence check of the correct assertion. The time of human action is not deterministic and it might require the user to find her hardware token, plug it in and execute the required action. 
An adversary can \fix{ensure the user is ready} by executing the time measurement attack on the second consecutive WebAuthn assertion so that the user is already prepared for a user presence check.
\fix{Further,} to maximize the measurable time difference, the adversary can introduce multiple repetitions of the same key handle in the \allCred parameter, \fix{spreading the user indeterminacy across multiple CTAP calls}. 
\fix{We outline a second approach} in Section~\ref{microphone-time} \fix{using the user's microphone to determine the point at which the user clicks the token's button and hence when key handle processing begins.} 

\subsection{Difference in Key Handle Processing}
\label{subsec:difference-in-key-handle-processing}

FIDO authenticators use unique keys per relying party. This ensures the unlinkability of the user's accounts, \fix{however also} introduces a storage problem since hardware tokens have limited memory and can only store a \fix{small number} of on-device keys (also called resident keys). \fix{A} common cryptographic technique used as a solution for this problem is key wrapping, i.e. storing \fix{an} encrypted \fix{version} of the secret key outside of the device.

Yubikey hardware tokens (manufactured by Yubico) are amongst the popular ones \fix{that} use this technique~\footnote{See e.g. how keys are stored in case of Yubico:  \url{https://developers.yubico.com/U2F/Protocol_details/Key_generation.html}}. 
Yubico's approach is to wrap the signing keys together with the corresponding origin/application ID or its hash value (to have a constant size plaintext instead of a variable one). This ensures that the key handle can only be used with the correct relying party. It also protects against simple linking attacks where two malicious relying parties can try sending key handles from the other party to identify users.
To protect the integrity of the plaintext authenticated encryption (AE) is used. The standard approach is to use the encrypt-then-mac approach, i.e. compute a message authentication code (MAC) on the ciphertext. 

Key wrapping is usually done using a constant size master key. This is to limit the size of the key material stored which is also one of the goals of key wrapping. It follows that provided ciphertexts from different RP's the token will always correctly decrypt and verify the MAC. However, depending on the plaintext the actual execution can differ. In particular, after comparing the origin in the key handle with the one given as input, the token can either abort execution (in case of failure) or create an assertion (if the origin is accepted).

The simplest way one would implement this process is first to check the validity of the MAC and abort in case of failure, and then proceed with checking the origin. The former requires the computation of the hash value of the input origin. As noted above we have three cases: 1) abort on MAC verification, 2) abort on origin verification, and 3) complete execution.
If the above implementation is used then there should be a time difference between cases 1) and 2). 

We will now show that this is what probably happens for the hardware tokens for which we were able to prove the existence of a timing difference. It is worth noting, that without the firmware of the vulnerable tokens we are unable to pinpoint the actual reason for the difference. 

To give an argument supporting our proposition let us take a look at the key wrapping decryption process implemented in Google's open-source OpenSK FIDO token implementation \cite{opensk}. We \fix{show} the interesting part of the OpenSK source in Figure~\ref{fig:opensk} (Appendix \ref{apdix:implementation}). In particular lines 272 and 295. In the former, the token verifies the MAC for the key handle and aborts in case it is invalid. In the latter, the token compares the decrypted id of the relying party with the origin provided as part of the FIDO authentication data to ensure that the key handle is only used with the correct relying party. Between those lines of code (lines 279-294) the token is doing other computations that, among others include AES CBC decryption of the key handle. It is easy to see that depending on the computational capabilities of the hardware this can lead to a noticeable time difference.

\fix{\subsection{Adversarial model}}
\todo{3. [C,D] Related to the adversarial model, the reviewers wondered about the attack in practice. The revision should make more clear the practical significance and impact of the attacks, alongside some sense of the prerequisites required for this attack.}

\fix{The goal of the adversary is to link FIDO2 registrations that were executed using the same authenticator. Under the assumption that the same authenticator is used by the same person, the adversary then has a connection between that user's accounts, and effectively create a user profile. 

We assume the adversary has the capability to remotely control the flow of FIDO2 authentication, however the attack does not deviate from the FIDO2  specifications. }
Our adversary is an active attacker \fixfix{in the sense that they control JavaScript code executed on the victim's client and manipulate FIDO communications, however,} the attack can only be performed \fix{during an authentication} transaction initiated by the victim\fixfix{, and }\fixfix{can only} \fix{leverage valid modifications of FIDO2 messages without disrupting the protocol or deviating from the protocol definitions (workflow, syntax, validations). }

\fix{
To achieve these capabilities, the adversary either needs to be a trusted authentication provider or have the ability to inject malicious code and payloads into the authentication process (see Section~\ref{sec:possible_adv} for examples). In this work we exclude adversaries that can execute CTAP calls directly (e.g.,~a compromised browser or malicious FIDO client). Adversaries with this stronger capability can directly execute silent authentications both to determine if key handles are present on the authenticator (user account linking) and to maliciously authenticate without user knowledge. 

Below, we summarize attacker capabilities.\\
Adversary can:
\begin{itemize}
    \item execute and manipulate FIDO2 protocol messages,
    \item access key handles (owned or stolen) not presented by the victim,
    \item measure timing of WebAuthn authentication calls.
\end{itemize}

Adversary cannot:
\begin{itemize}
    \item deviate from FIDO2 protocol specifications,
    \item trigger errors (as these would alert the victim).
\end{itemize}
}

\subsection{Attack concept}
First, notice that 
\fix{the adversary} described above is in possession of data that includes key handles corresponding to the authenticators of users. \fix{For simplicity we} focus on a single \fix{attack} scenario: 
an adversary that implements service $\mA$ and tries to distinguish if the key handle $\khandle_\mB$ for service $\mB$ corresponds to the user \fix{authenticating} with handle $\khandle_\mA$. The malicious queries can be build out of $\khandle_\mA$, $\khandle_\mB$ and random handles $\khandle_\mR$. We \fix{use} random key handles as \fix{a proxy for} key handles generated by different authenticators. 
The attack is successful assuming there is a noticeable difference in the time it takes the authenticator to process key handle $\khandle_\mB$ and $\khandle_\mR$ when connecting to service $\mA$. We show \fix{in Section~\ref{subsec:hardware_authenticators}} that this assumption is actually true for some existing hardware tokens \fix{and in Section~\ref{subsec:difference-in-key-handle-processing} we discuss} potential reasons for this time difference.


\begin{figure}[htp]
    \centering
    \includegraphics[width=\columnwidth]{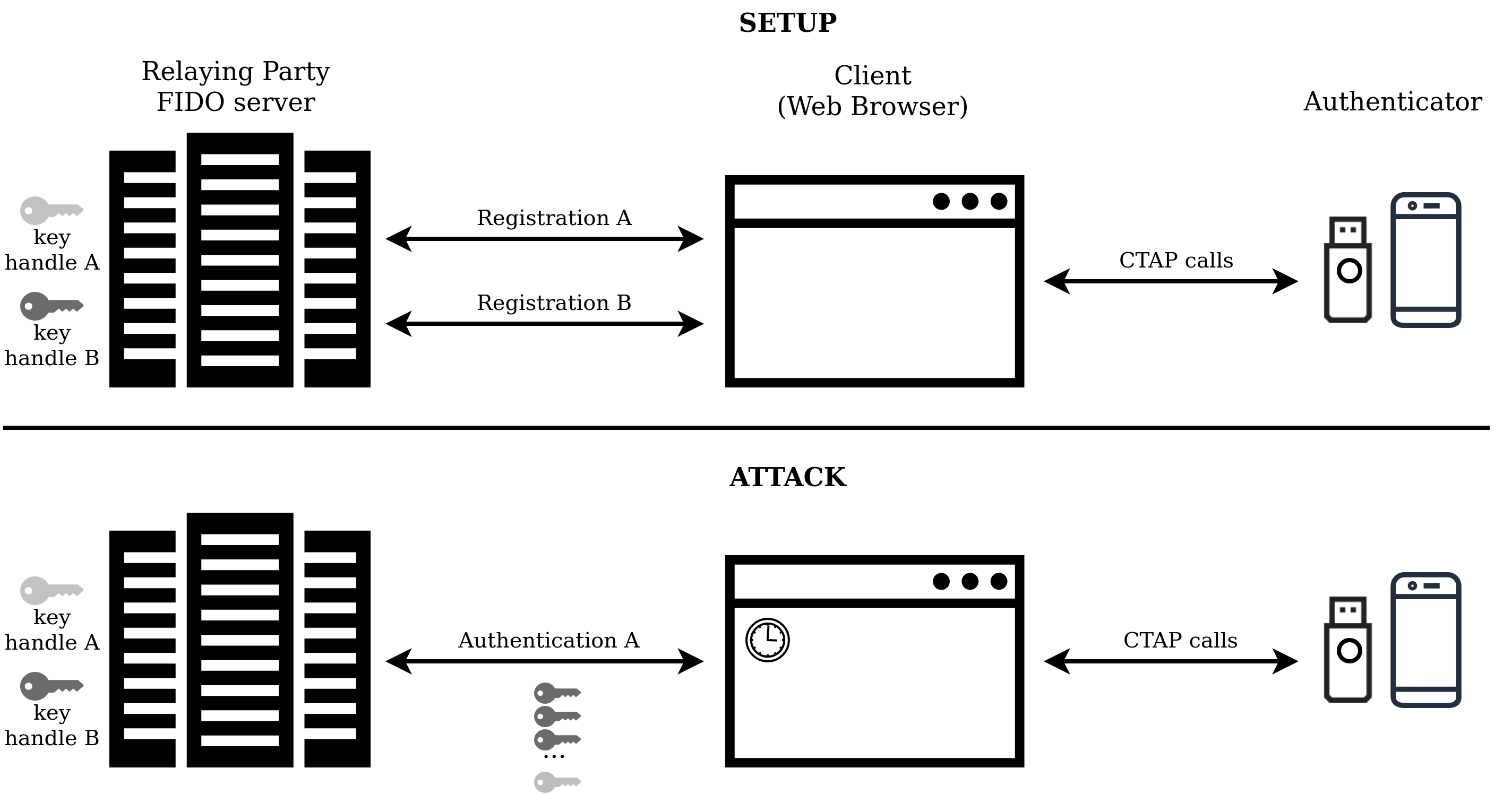}
    \caption{\fix{During authentication to $\mA$, the attacker builds an \allCred list with multiple $\khandle_\mB$ and $\khandle_\mA$}.  }
    \label{fig:attack}
\end{figure}

\begin{figure*}[ht!]
    \centering
    \includegraphics[width=\textwidth]{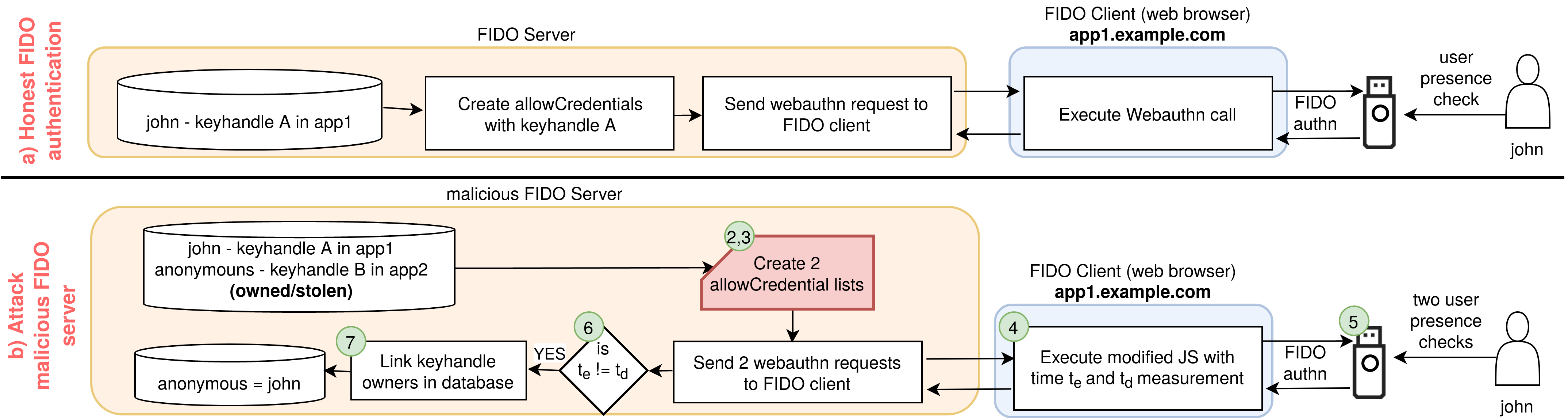}
    \caption{\fix{FIDO2 timing attack example diagram. \fixfix{Numbers in green circles correspond to steps in Algorithm \ref{alg1}.} More attack examples can be found in Appendix \ref{appx:attack_scenarios}.\\
    a) The first diagram illustrates an honest FIDO transaction.\\
    b) The second diagram presents an attack flow triggered by a malicious FIDO server.}}
    \label{fig:attack_simple_example}
\end{figure*}

The first step of the attack is to find the baseline execution time $t_e$, which corresponds to the time it takes for an authenticator to answer a \fix{WebAuthn} API call with $n$ random key handles and one valid key handle for service $\mA$ plus the time for the non-deterministic user factor. 
\fix{The key handle list is placed in the \allCred field of the call (see Figure~\ref{fig:attack}).} 
To measure the time the adversary uses timers to enclose the {\sf navigator.credentials.get} call to the browsers \fix{WebAuthn} API.
The \fix{second part of the} attack is performed in a similar manner, however this time \fix{$n$ copies of $\khandle_\mB$ are used in place of random key handles.} 
The resulting time $t_d$ is then compared with $t_e$. In case of a noticeable difference the adversary concludes that $\khandle_\mB$ is also registered in the authenticator, and the user has an account with service $\mB$.
Note that once $t_e$ is know the adversary can omit the first step of the attack (assuming the same user for $\mA$ is connecting). 
\fix{Algorithm \ref{alg1} provides an overview of the attack.}

The value $n$ (the number of requested silent \fix{authentications}) is an \fix{attack} parameter and depends on the attacked platform. The higher it is the less the non-deterministic user factor influences the attack, since by replicating the key handle in question (\fix{e.g.~}$\khandle_\mB$) \fix{it is} divided across all executions.

\begin{algorithm}[H]

\begin{algorithmic}[1]
\State $Victim$ : starts FIDO2 transaction
\State $Adversary$ : prepares \allCred list with n random $\khandles$ and $\khandle_\mA$ (can be omitted if $t_e$ known)
\State $Adversary$ : prepares \allCred list with n copies of $\khandle_\mB$ and one $\khandle_\mA$
\State $Adversary$ : forces FIDO client to perform  WebAuthn calls with prepared \allCred lists and measures times of execution ($t_e$ and $t_d$)
\State $Victim$ : performs user presence checks
\State $Adversary$ : compares measured times ($t_e$ and $t_d$)
\State $Adversary$ : Links identities if times do not match
\end{algorithmic}
\caption{High level attack algorithm}
\label{alg1}
\end{algorithm} 

\fix{
\subsection{Possible adversaries}\label{sec:possible_adv}
}

\fix{We present five possible adversary types: three that are providers of FIDO2 authentication, one that is capable of intercepting and manipulating FIDO2 and client side JavaScript}\fix{, and one that uses injected JavaScript code to trigger the attack. }

\fix{All adversaries execute the same attack concept, however details of the attack setup differ. }
\fix{The two \allCred lists (one containing random key handles, the other a candidate key handle ---  Algorithm~\ref{alg1} steps 2,3) may be created on the FIDO server (as shown in Figure~\ref{fig:attack_simple_example}) and communicated to the client via WebAuthn transactions, may be created in a malicious server and inserted into WebAuthn transactions, or may be created by attacker JavaScript code on the client. 
In all cases JavaScript code timing the execution of WebAuthn calls is then executed on the client (Algorithm~\ref{alg1} step 4), the user performs user presence checks (step 5) and results sent to the attackers server (steps 6,7). 
Schematics of these additional attack scenarios are illustrated and described in Appendix \ref{appx:attack_scenarios}. }

The first \fix{adversary type provides FIDO services for a single application} 
\fix{where} users can benefit from multiple accounts. 
\fix{An example of} this use case \fix{is} a cryptocurrency exchange (e.g.,~one of the biggest cryptocurrency exchanges, Binance, implements FIDO2 as a second factor). Having multiple accounts \fix{on the exchange} can reduce the traceability of one's transactions, hence keeping them unlinked is of great value for the user. 

The second \fix{consists of} more complex applications that provide not only core services but can also \fix{act as an identity provider, allowing \fix{users} to login to third party applications using their accounts and the login flow of the identity provider. }
Examples are services similar to Google which provide a "Login with XYZ" service. In this setup, we can imagine users \fix{with two Google accounts} that do not want \fix{associated third party accounts to be linked.} 

The next group offer FIDO2 as a service for third part\fix{ies}. They make it easy to integrate strong authentication but it also means that FIDO2 related data is kept on their \fix{servers and} they execute FIDO2 authentication flows. An example \fix{provider} of such a service is Duo Security. \fix{Such} service providers are in possession of  data from different services \fix{and hence} are capable of executing cross service linking.

\fix{Our last two possible adversaries are not owners of FIDO2 data, but can obtain either stolen data or spy on user authentications. }
\fix{First,} any service acting as an SSL termination proxy can intercept and modify FIDO2 message payloads {and modifying FIDO related javascript,} \fix{and are thus} capable of \fix{sending} malicious payloads \fix{and executing timing code on the client}. \fix{A} commercial example of such a service is Cloudflare. Similarly to the FIDO2 as a service case, the proxy can gather FIDO2 data from \fix{a diversity of} applications. 

\fix{Finally, any actor that is able to modify JavaScript code is in a position to execute our timing attack. Considering that \fix{the} JavaScript XSS (Cross Site Scripting) attack vector \fix{has remained} in the OWASP TOP 10 list of vulnerabilities \fix{for many years} \cite{OWASP_top10}, we \fix{consider} this variation of the attack as highly probable. Similarly to the proxy example, FIDO2 data can be gathered from user authentications in compromised browsers or obtained from data breaches. Note that FIDO2 credentials \emph{cannot} be stolen in this way.}

\fix{We note that in all cases, stolen FIDO2 data (in the form of key handles) can be utilised by attackers to broaden the attack scope.} \fixfix{Additionally, adversaries can use context metadata (e.g. account data linked to key handle) to narrow down the set of potential key handles to check.}

Considering the number of possible adversary types, we believe that the attack described below has a high potential to be deployed in the real world and can violate the privacy of users secured with FIDO2.

\todo{ [A,B,C] The reviewers also wanted to understand the attack layout better, such as through a fully worked example of how all of the pieces fit together. There are many ways this can be accomplished, potentially involving a worked example (noting the attack model) and/or a diagram.}

\section{Results} \label{results}

In this section we present the methodology and results of our tests. FIDO2 with WebAuthn is designed for web applications and it specifies a well-defined path between the relying party, browser, and authenticator (as shown in \fix{F}igure \ref{fig:flow}). 
Our attack faithfully follows the FIDO2 flow, which executes through FIDO clients and authenticators. 
For this reason we focus our analysis on those two elements. Our test sessions were recorded and are available online together with source code\footnote{Test recordings: \url{https://osf.io/t7dpa/?view_only=c8595da6c6d34fadb87f2f6db7e5d626}}.

\begin{table*}[htp!]
\center
\renewcommand{\arraystretch}{1.2}
\caption{Test results indicating hardware tokens vulnerable to timing attack.}
\begin{tabular}{p{2cm}||x{1.3cm}|x{1.6cm}|x{1.3cm}|x{1.4cm}|x{1.3cm}|x{1.4cm}|x{1.6cm}|x{1.6cm}}
  & \textbf{Yubikey 5} & \textbf{Hyperfido Titanium Pro} & \textbf{Google Titan} & \textbf{Token2 T2F2-Bio} & \textbf{Feitian K26} & \textbf{TrustKey G320H} & \textbf{AuthenTrend ATKey.Pro} & \textbf{Kensington Verimark Guard}
  \\ \hline\hline
 Is vulnerable& \xmark & \cmark (10.07)$^*$ & \xmark & \xmark & \cmark (2.21)$^*$ & \xmark & \xmark & \xmark  \\ \hline
  Is upgradeable& \xmark & \xmark & \xmark & \xmark & \xmark & \xmark & \xmark & \xmark
\\ \hline
\multicolumn{7}{l}{\scriptsize $^*$ Average difference (ms) for silent authentication between random and bad origin key handles} \\

\end{tabular}

\label{table:hardware_results}

\end{table*}

\subsection{Methodology}
The methodology of our test suite includes two parts. Firstly, we measure silent authentication directly on the  authenticators to identify vulnerable devices. In the second part, we executed remote timing measurements on the WebAuthn API using the vulnerable devices from the first phase. 

In the first phase, our goal was to measure silent authentication directly on the FIDO2 authenticators. 
For the USB hardware authenticators we used the open source Yubico FIDO2 library to make CTAP calls directly. Unfortunately, the same is not possible on \fix{the} Android platform because the available SDK does not implement direct access \fix{to} CTAP and the WebAuthn API forces a user presence check. 
In the case of iOS, access to both WebAuthn and CTAP are unavailable, and we were unable easily to test for time differences. 

We measured the time between request and response for multiple independent silent authentication calls with a single key handle in the \allCred list. We used either a random key handle (with the correct length) or a correct key handle with an incorrect origin value. When there was an appreciable difference, we identified the authenticator as vulnerable. 


For the second phase we built a Relying Party in Node.js which serves a test HTML page with WebAuthn API executions in JavaScript and measures execution times. 
We deployed the software on Amazon AWS 
with a public IP address which we accessed via several web browsing platforms to obtain authentication timing data. 


\subsection{FIDO2 Hardware Authenticators}
\label{subsec:hardware_authenticators}
The number of authenticators available on the market is constantly growing. Because FIDO2 is an open source specification, there is no public record of commercial FIDO2 authenticators. However, an estimate can be made based on FIDO Alliance voluntary certification, which at the time of this writing holds 140 certified FIDO2 authenticators. The list contains platform, roaming, hardware and software authenticators. Notably, our attack can be launched against any type of FIDO authenticator. Due to time restrictions we chose to evaluate the most secure and numerous (111 certified devices) FIDO authenticator type:  hardware roaming authenticators. 

We selected eight hardware authenticators that are certified by the FIDO Alliance at level 1 \cite{fido_lvl_1} with an aim to provide a representative view of the available options. \fixfix{Our selection provides a broad range of price ranges, vendor sizes, features, and countries.} We picked ``Yubico Yubikey 5 FIDO2 USB-A'', ``HyperFIDO Titanium Pro'', ``Google Titan'', ``Token2 T2F2 Bio'', ``Feitian K26'',  ``TrustKey G320H'',  ``Kensington Verimark Guard'', and  ``AuthenTrend ATKey.Pro''. 

While examining the Yubikey token, we observed a defense mechanism: 
After 10 incorrect attempts a randomised delay is added,  which prevents our attack from succeeding. 
Google Titan, Token2 T2F2, TrustKey G320H, Kensington Verimark Guard, and AuthenTrend ATKey.Pro \getAssert times were not distinguishable. 
We successfully executed timing attacks on HyperFIDO Titanium Pro (average difference of 10ms per execution) and Feitian K26 token (average difference of 2ms per execution) --- see Table \ref{table:hardware_results} for an overview and 
Figure~\ref{fig:hyper_times}, left panes for timing results for vulnerable tokens (see Appendix~\ref{Apdix:silent-auth-times} for other tokens). 

A simple threshold based classifier correctly identifies key handles with a 0.1\% error (HyperFido) and 6\% error (Feitian) if user presence timing is known (for example, using an audio signal as described in Section~\ref{microphone-time}). We further simulated noise from user presence checks using results from a small user study (see Section~\ref{subsec:user_study}) by adding a randomly sampled user presence check timing result to each CTAP timing result. These adjusted CTAP timing figures (one set of figures for each subject in the user study) were then used to determine a threshold as before and to predict whether each CTAP call contained key handles present on the token (
Figure~\ref{fig:hyper_times}, right panes). In all cases, 70\% of the CTAP timing data was used to determine the threshold and the remaining 30\% to evaluate the resulting classifier. 
Note that an attacker may combine estimates from multiple user authentication sessions to further mitigate noise from user presence checks. 



\begin{figure*}[h]
    \centering
    \includegraphics[width=\textwidth]{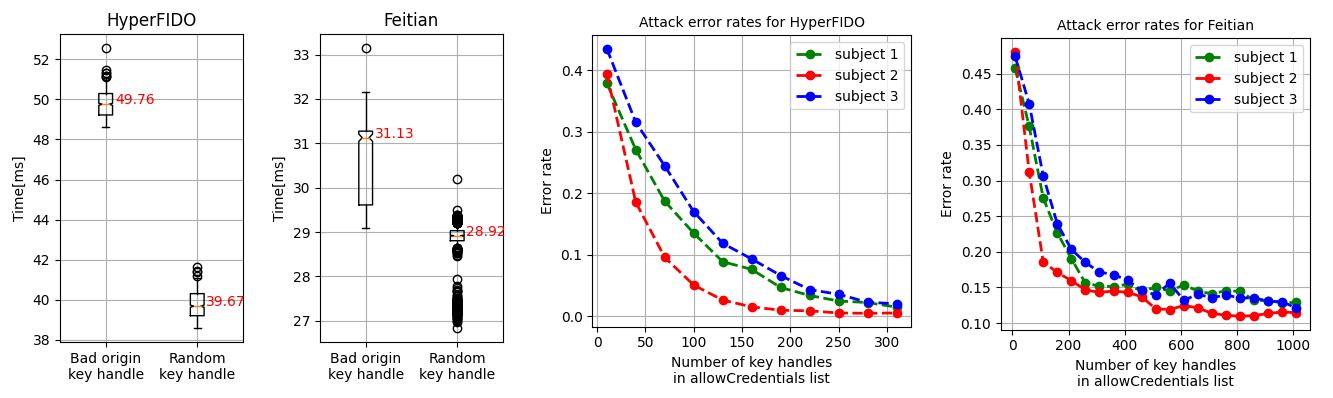}
    \caption{Measurements of response times of vulnerable tokens HyperFIDO Titanum PRO and Feitian (left). Right: Error rate for a simple threshold classifier with user presence noise from the user study.}
    \label{fig:hyper_times}
\end{figure*}

\subsection{FIDO2 Clients}
The second element in the FIDO2 flow that \fix{can} be vulnerable to timing measurements of assertions is the FIDO2 client. The most popular FIDO2 clients are web browsers. If the WebAuthn API is supported by the browser, each execution of the WebAuthn API in JavaScript is translated into CTAP calls to the FIDO2 authenticator. 

We tested six popular web browsers (we included the ``Brave'' browser as the one that focuses on privacy) running on 5 widely used operating systems for desktop and mobile (see Table~\ref{table:OS_CTAP}). We did not evaluated Internet Explorer because it does not implement WebAuthn. We used the latest versions of browsers as of 6th of April 2021. Four of \fix{the} six tested browsers are based on the Chromium engine (only Firefox and Safari have completely independent source code). 
We used a HyperFIDO Titanium Pro hardware token because we knew that it is vulnerable to timing attacks (Section~\ref{subsec:hardware_authenticators}). 

On the Windows platform (Windows 10 Home 19042.867) \fix{all supported browsers passed control to}  
the Windows WebAuthn API. 
We were unable to execute the attack with a \fix{large} number of key handles. The Windows WebAuthn API introduces a limit on how many requests can be sent to the token. Empirically, using Wireshark USB logs we confirmed that 20 silent authentication attempts are made before failure.

On MacOS Big Sur (Version 11.2.3 on MacBook Pro), all tested browsers showed vulnerability for our timing measurement. We experienced unexpected behavior from the Safari browser: 
Test attempts with 64 or more key handles in the \allCred list cause Safari to crash, hence our attack was limited to only 63 key handles. 
Though this would reduce the efficiency of our attack, we recognize it as a bug and not a security feature, thus we conclude that Safari is vulnerable. 


The last desktop platform which we tested is Ubuntu 18.04 (one of the most popular desktop operating system from the Linux family) for which we were able to successfully execute timing measurements on all browsers. 

We tested Android 10 as it is the most popular Android OS version at the time of this writing. The test was executed on five phones: Google Pixel2, Samsung A2, Mi8, Motorola One Vision and Oppo Reno2 Z. All tested phones are equipped with a fingerprint scanner. We found that only Chrome and Opera support WebAuthn executions. Similarly to Windows, WebAuthn control is given to the Android system where the user can select which token type should be used. We tested a native Android authenticator secured with fingerprint (the "Use this device with screen lock" option). We did not observe a timing difference during our attack. 

The iOS platform has the most limited group of browsers supporting FIDO2 \fix{as} Apple has not yet released a native API for WebAuthn. In our tests (executed on iOS 14.4.2), only Safari and Brave allowed WebAuthn calls. Safari uses iOS native APIs which allow using iOS authentication mechanisms (\fix{e.g.,} TouchID). We \fix{found} that the iOS API works as a client with client-side storage and because we couldn't see any difference in response times, we suspect that the \allCred list is filtered with key handles saved on the client-side. The Brave browser uses a custom implementation to connect to a hardware FIDO2 token. We were unable to check \fix{our} physical tokens on iOS because of the incompatibility of \fix{the} iOS lightning port with our tokens. 

\begin{table}[htp!]

\center
\renewcommand{\arraystretch}{1.2}
\caption{Test results indicating if browsers execute silent authentications for all key handles in \allCred list, thus allowing timing attacks when combined with a vulnerable authenticator.}
\begin{tabular}{l||c|c|c|c|c|c}
  & \rotatebox[origin=c]{90}{\textbf{Chrome}}
  & \rotatebox[origin=c]{90}{\textbf{Brave}}
  & \rotatebox[origin=c]{90}{\textbf{Firefox}}
  & \rotatebox[origin=c]{90}{\textbf{Opera}}
  & \rotatebox[origin=c]{90}{\textbf{Edge}}
  & \rotatebox[origin=c]{90}{\textbf{Safari}}
  \\ \hline \hline
Windows 10$^*$ & \xmark & \xmark & \xmark & \xmark & \xmark & N/A
\\ \hline
MacOS 11.2.3   & \cmark & \cmark  & \cmark & \cmark  & \cmark  & \cmark
\\ \hline
Ubuntu 18.04 & \cmark & \cmark & \cmark & \cmark & \cmark & N/A 
\\ \hline
Android 10        & \cmark           & N/S & N/S & \cmark           & N/S & N/A            
\\ \hline
iOS 14.4.2 & N/S & - $^\dagger$ & N/S & N/S & N/S & \xmark  $^{\ddagger}$ \\ \hline 
\multicolumn{7}{l}{\scriptsize \cmark, \xmark$\,$- Allows / does not allow attack}\\
\multicolumn{7}{l}{\scriptsize N/S - FIDO2 not supported}\\
\multicolumn{7}{l}{\scriptsize $^*$ Browsers use native Microsoft WebAuthn API}\\
\multicolumn{7}{L{.4\textwidth}}{\scriptsize $^{\dagger}$ Brave browser for iOS has a custom implementation that uses hardware tokens only}\\
\multicolumn{7}{l}{\scriptsize $^{\ddagger}$ Safari uses native iOS WebAuthn API}\\

\end{tabular}

\label{table:OS_CTAP}
\end{table}

%

\subsection{Dealing with User Presence Checks}
We have seen that using multiple key handles in a single CTAP call can reduce the impact of the indeterminacy of the time taken by the user to perform the user presence check. In this section we present present two additional approaches to reduce it's impact and a small pilot study to quantify that indeterminacy. 

\subsubsection{Priming User Presence Checks}\label{subsec:user_study}
\todo{6. [Reviewer discussion] There was concern raised in the discussion among reviewers about the ethics and appropriateness of the small-scale study reported. The reviewers are hoping that the revision clarifies the study population (seemingly the authors themselves?) and any relevant discussions with their IRB. An experiment on the authors themselves may not require IRB approval. The reviewers are simply hoping that the revision more accurately describes what was done, and why.}
The first strategy to reduce the impact of user presence checks is to prime the user by requiring them to perform the check twice: the first time intended for them to find the token and insert it, the second for timing. The user would be told that there was a problem with authentication and that they need to repeat it. Our hypothesis is that this approach would lead to substantially more consistent timing for the second check. 

To verify this hypothesis we performed a small \fixfix{proof of concept} study \fix{among authors} that simulated this sequence of events. 
We built a simple web page that requests FIDO authentication. 
Our subjects were instructed to insert a FIDO2 hardware token once the browser shows the authentication prompt and take it out once authentication is successful. We notified them that authentication might not work first time, and in that case the token doesn't need to be removed between attempts. The authentication was repeated 50 times for each subject. Each authentication was triggered after a randomized time interval \fixfix{to limit preparedness and thus minimise bias}. The results show that the second consecutive authentication takes far less time and has far less variability as we hypothesised \fix{(see Table \ref{tab:user_study}). All study participants were authors, and thus the study was exempt from ethics review (confirmed by IRB). }
\fixfix{We acknowledge that, despite our best intentions and measures to minimise bias, the study was in the end conducted by authors and bias may remain.}

\begin{table}[]
    \centering
    \caption{Timing variation results for FIDO2 authentication time from user study. }
    \label{tab:user_study}
    \begin{tabular}{c|m{1.4cm}|m{1.4cm}|m{1.4cm}|m{1.4cm}}
        & \multicolumn{2}{c|}{1st authn} & \multicolumn{2}{c}{2nd (primed) authn} \\
        \hline
        Subject & Mean & Std. Dev. & Mean & Std. Dev. \\ \hline
        1 & 5041 ms & 943 ms & 750 ms&	227 ms\\
        \hline
        2  & 3980 ms & 585 ms & 344 ms & 116 ms\\
        \hline
        3  & 5441 ms &  844 ms & 707 ms &	277 ms	 \\
    \end{tabular}
\end{table}

\begin{figure*}[htp!]
    \centering
    \includegraphics[height=4cm]{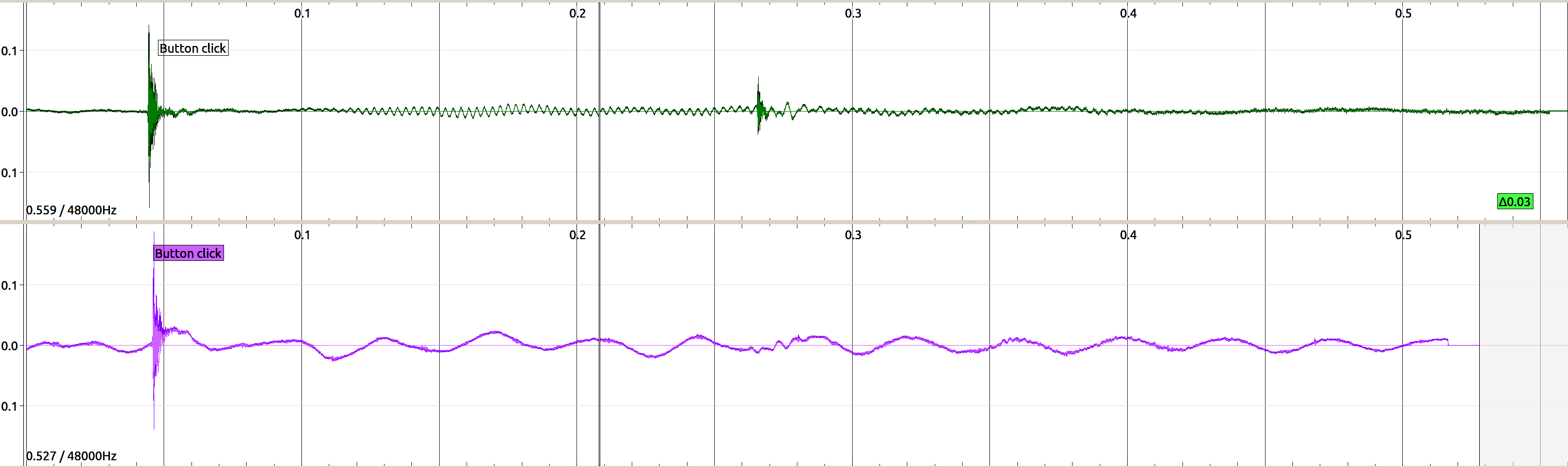}. 
    \caption{Audio recording of FIDO assertion on HyperFido Titanium Pro. Green: an attempt with a bad origin key handle. Pink: an attempt with a random key handle. The initial silence is truncated to allow comparison.}
    \label{fig:mic}
\end{figure*}

\subsubsection{Alternative Time Measurement} \label{microphone-time}
In our attack \fix{the} adversary builds an \allCred list \fix{with multiple} instances of the same key handle to limit the influence of user action. Some configurations are resistant to this kind of attack by limiting the number of allowed entries in the \allCred list \fix{(see Section~\ref{results})}. To circumvent that limitation we propose an alternative method of measuring the execution time \fix{in which} the measurement starts \fix{immediately} after the user presence check. This way we eliminate the non-deterministic delay and can use smaller sized \allCred lists.
%

The \getAssert WebAuthn API call implemented by browsers enforces a user presence check which is not uniformly implemented in different FIDO2 authenticators. 
%
This check requirement introduced additional elements in the manufacturing process of hardware tokens. The factors that influence the decision on what kind of interface is selected include not only security but also production cost and user experience. For example, some Yubico tokens use capacitive touch sensors whereas HyperFido tokens use a physical button, which is one of the most popular solutions. Inspired by the work in the area of acoustic side-channel attacks (e.g. Genkin at al. \cite{8835386}) we observed that physical buttons \fix{on authenticator tokens} emit a characteristic sound when used. We use this observation in our modified attack described below.


\subsubsection{The Modified Attack}
In this variation of the attack, the service does not have to send multiple keys, however it needs to record audio using the attacked platforms microphone. This can be achieved using the MediaStream API \cite{w3c_mediacapture}, which allows \fix{the} capture \fix{of} sound directly to a JavaScript object. We wrapped the execution of \textit{navigator.credentials.get} with sound recording code. After the recording is finished, the sample is sent to the backend part of the attacker/service for processing. Figure \ref{fig:mic} presents recordings from two attempts, using a key handle with bad origin and using a random key handle. The test was performed in a home environment on a Dell XPS 15 9570 laptop with HyperFido Titan Pro token. The button click action is easily distinguishable and the time difference between attempts simplifies the identification of the key handle with bad origin.

It is easy to see that this attack requires a strong adversary that is granted access to the microphone through the MediaStream API. It also requires that the time difference between executions of random and bad origin key handles are high enough to be distinguishable in the recording, \fix{and} the token needs to be constructed with a button that generates noticeable sound.  
\fix{Many services use the} MediaStream API to provide videoconferencing features and once consent is given, the application can trigger recordings freely and would be in position to perform the attack. 
Interestingly, the microphone usage consent window is presented to the user as a standard browser dialogue window which appears with the same location and ``look and feel'' as WebAuthn dialogues, and is shown just before the WebAuthn window, so it could be easily accepted by mistake, \fix{enabling the attack in other cases}.

\subsection{User Experience} Our attack has minimal influence on how FIDO2 authentication is perceived from a user perspective. 
Authentication proceeds as normal due to the correct key handle at the end of the list, but with a slight delay. 
The time difference can be noticeable but it is indistinguishable from network or browser slowdown. Therefore, we claim that it is unlikely for the user to notice any irregularities in the authentication process. Examples of user experience can be observed in attached recordings.


\subsection{FIDO in the Wild}
\label{sec:fido-in-the-wild}
In this section we discuss how FIDO2 is deployed in publicly available web applications and how this relates to our proposed attack. In terms of production applications, security of the solution is not the only aspect to be considered. For example, introduction of an additional factor for authentication brings additional cost and potential disruption to the business. Perhaps the most surprising aspect of the FIDO2 environment is user adoption and experience. From past studies about the usability of FIDO2 hardware tokens~\cite{9152694}\cite{238317} we understand that the transition to more secure ways of authentication is facing challenges of a human nature.


Our proposed timing attack is based on a non-resident key scenario, which we believe covers the majority of publicly available FIDO2 implementations. To the best of our knowledge there is no previous research that attempts to quantify FIDO2 in the wild. To better understand the scale of applicability of our timing attack on FIDO2 implementations we developed and executed a web crawler. We checked the 1 million most popular DNS records from the Cisco Umbrella set for the presence of WebAuthn protocol executions between 10 December 2020 and 20 December 2020. Because WebAuthn executions can be found in the javascript resources of web applications we targeted our crawler to check for the presence of the \textit{navigator.credentials.create} property with \textit{public-key} type. We acknowledge that some applications have more complex authentication procedures that do not reveal usage of the WebAuthn protocol (\fix{e.g.,} WebAuthn is dynamically loaded after completing first factor of authentication). Fortunately for us, initial authentication (\fix{e.g.,} username/password) implies that FIDO2, if used, will be configured as a second factor without resident keys \fix{in these cases}. 

The results confirmed our belief about FIDO2 usage in the wild. We gathered 684 records of WebAuthn executions from which we extracted following groups. Most findings (52\%) came from applications that reuse open-source tools (\fix{e.g.,} the discord platform is frequently used as community forum, nextcloud is used as a file storage and share service). In those cases FIDO2 is implemented \fix{with non-resident keys (as second factor authentication)}. The second identified group (16\%) consists of federated authentication platforms (\fix{e.g.,} wordpress.com, github.com, microsoft login). In this group only Microsoft login \fix{with passwordless FIDO2 used resident keys}. In the last group, we gathered applications that use self-implemented FIDO2 authentication, from which all were configured to use FIDO2 \fix{with non-resident keys (as a second factor)}.

Our final conclusion from the crawling exercise is that in the public space FIDO2 is mostly used as a second factor mechanism with non-resident keys. Passwordless authentication (with resident keys) is in the early adoption phase and only a few providers give this option. In terms of our timing attack, this means that the majority of Relying Parties in the wild \fix{are vulnerable to and} 
have the potential to launch our attack to link key handles. 

\section{Related Work}
The FIDO2 standard only recently gained increased attention from the research community even though it was introduced in 2016, \fix{perhaps due to} the increased adoption and popularity of FIDO2 in recent years. 

Barbosa et al. \cite{cryptoeprint:2020:756} introduced a formal model for authentication based on WebAuthn and CTAP. They showed that FIDO2 is secure in the model but also proposed an improvement that fulfills stronger notions of security. Guirat et al. \cite{10.1145/3190619.3190640} performed an analysis of the WebAuthn protocol, which revealed that formally some authenticators might leak privacy. Research done by Feng et al. \cite{UAFAnalysis2021} gave insights on the UAF protocol, showing its unlinkability property. Lomne et al. \cite{titan_ninja} successfully extracted key material from a Google Titan token and were able to clone it. A different view was presented by Alam et al. \cite{10.1145/3319535.3363283}, who address potential security and privacy issues caused by development flaws.

Klieme et. al. \cite{9343231} proposed a solution for continuous authentication of the user using silent authentications and FIDO2 extensions. Chakraborty et al. \cite{10.1145/3319535.3363258} proposed the use of a sim card TPM implementation (simTPM) as a secure and convenient FIDO2 authenticator. Dauterman et al. \cite{8835225} introduces an enhanced design for a token, which is resistant to backdoor attacks and prevents privacy loss by token fingerprinting. Frymann et al. \cite{10.1145/3372297.3417292} analyzed Yubico's proposal for backup tokens including verification of unlinkability of paired tokens.

Lyastani et al. \cite{9152694} present extensive results on usability and acceptability of FIDO2. Surprisingly, users' privacy concerns did not influence the acceptability rate. Pfeffer et al. \cite{272198} shed light on the usability of authenticity checks in the case of physical tokens. Florian et al. \cite{255646} present a usability study of FIDO2 authentication in a small company. The results revealed that despite security benefits, hardware tokens face acceptability challenges.

\section{Discussion}
\fix{
Here, we discuss the features that were key enablers of the attack presented in this paper. We hope that our findings will contribute to enhanced security of all certified FIDO2 tokens. 

Firstly, the silent authentication mechanism opens a path to bypass the human factor in FIDO2 authentication. We acknowledge that it simplifies the automation of the pre-authentication processes, nevertheless it introduces threats that, from a privacy perspective, may outweigh its benefits. 
We believe that FIDO client's implementations should introduce additional safeguards on the silent authentication process (as described in Section \ref{sec:attack_mitigation}).
Moreover, we observed that rate limiting techniques (e.g., adding delay after a number of unsuccessful calls) are not popular in FIDO2 authenticators. This might allow for enumeration, brute-force or as in our case timing attacks. 

Additionally, we want to emphasize the importance of FIDO Alliance certification. We acknowledge the utility of graded certification levels, however understanding differences between certification levels requires expert knowledge not available to a regular user. Our attack showed that authenticators with L1 certification cannot guarantee all FIDO2 security and privacy goals. Therefore, we believe that all authenticators should be evaluated against L3 controls, which guarantee on-device testing. 
 }



\subsection{Attack Mitigation}\label{sec:attack_mitigation}
\todo{5. [B] The reviewers were hoping for at least a discussion of countermeasures in a bit more depth.

From B: \\
3) The authors mention that only the passwordless implementation used by Microsoft could not be deploying the attacks. What’s the reason about this fact? And does this have any enlightenment for the defense method proposed in this paper? It would be better if the author could add a discussion on this in the countermeasures part.}

\fix{The vulnerabilities responsible for our timing attack occur on two loosely coupled layers of FIDO2 transactions: FIDO clients and FIDO authenticators. Considering the number of authenticators and clients available on the market and already deployed, a complete mitigation solution has to address both layers. 

We present four mitigation strategies, two that apply to authenticators, one that circumvents our attack (at a cost) via FIDO configuration and one that applies to FIDO clients (e.g.,~browsers). }

\fix{\subsubsection{via Constant Time Execution in Authenticators}

The easiest way to protect against our attacks would be to update the firmware of hardware tokens and implement the execution in a way that there is no time difference between checking random key handles and key handles for different origins. \fixfix{In the sub-sections} below we  show how this can be done using existing techniques already implemented in some hardware tokens. 
Unfortunately, this only mitigates the problem in case of new users that buy a hardware token with updated firmware, as the majority of hardware tokens do not allow firmware updates.
}


\fix{\subsubsection{via Key Derivation Function in Authenticators}}
An alternative way to generate a keypair is to use a key derivation function keyed with a master secret and seeded with data related to the relying party. This process requires the authenticator to only store one master secret key (i.e. AES key) which is then used to pseudorandomly derive the signing key for the FIDO authentication process. 

This is a well-known technique and has already been implemented by some tokens. However, due to their closed firmware, we are unable to verify how many tokens implement key generation in that way. A notable exception is the SoloKey FIDO token which comes with an open-source firmware \cite{solokeys} for which
we show the key generation function (see Figure~\ref{fig:solokeys} Appendix \ref{apdix:implementation}). 
The implementation also uses key handles which in this case are just random values. During the authentication process, 
this random value is the used as the \verb+data+ argument for the key generation function ( \verb+data2+ is left empty). 

In this approach the key handles are just random values, \fix{hence} the \fix{authenticator}'s processing time for every key handle is the same and the \fix{authenticator} is not vulnerable to our attack.

\fix{\subsubsection{via Resident Keys (FIDO configuration)}\label{sec:mit_resident_keys}
Our attack utilises a weakness in incorrect implementations of key handling (i.e.,~handling non-resident keys) in authenticators. Methods such as key wrapping were introduced to solve a memory problem on authenticators, which need to store unique signing keys for each relying party (for privacy reasons). 
Alternatively, the FIDO2 protocol can be configured to store keys directly on the authenticator (residence keys), which eliminates our attack vector. Even though, the introduction of resident keys on the FIDO server might seem trivial (setting \textit{requireResidentKey} flag in the registration request), the consequences for FIDO authenticator users are significant. Firstly, not all FIDO authenticators support resident keys. Moreover, the modification of key storage technique in an existing authentication system requires all users to perform the FIDO registration process once again. Finally, the storage capacity of key handles in roaming authenticators is limited (e.g.,~Yubico keys allow up to 25 keys), which noticeably reduces their utility. 
We found only one deployment of FIDO in the wild that uses resident keys  (Microsoft AzureAD passwordless login). 
All other implementations we found in the wild (Section~\ref{sec:fido-in-the-wild}) use non-resident keys. 
}

\subsubsection{via Client Update. }
\fix{The FIDO client attack vector can} be mitigated by changing the way browsers \fix{and other clients} implement the WebAuth API \allCred parameter. In particular, we propose the following mechanisms. 

1) Deduplication of the \allCred list before making CTAP calls, which would remove all repetitions of bad origin key handles and thus prevent amplification of the time difference.\footnote{The Chromium team acknowledged our finding as an information disclosure vulnerability and suggested deduplication as a mitigation.} 

2) Silent authentication errors can be delayed by a random value that is large enough to render the attack ineffective due to a high error rates. 

3) The size of \allCred can be limited to e.g. 10 or 20 elements. This should still preserve the functionality since most users will never register more than 10 tokens with a single relying party but users remain vulnerable to the second (weaker) form of attack \fix{using audio detection of user presence checks}. Note that this is already the case with the Windows 10 WebAuthn implementation, which limits to 20 elements. 
%

Note that these approaches are independent of the hardware token and users can easily protect themselves by updating their browser to the latest version.

\section{Conclusion}

In this paper we introduced a conceptual attack against the privacy of the FIDO authentication process. At first glance the attack is based on strong assumptions about the capabilities of the adversary. However, we demonstrate   
that the chain of flows in protocol and existing implementations allows a remote adversary to break the unlinkability of FIDO2. We built a 
proof-of-concept and showed that the attack is possible for many configurations of FIDO clients and authenticators and showed that the majority of FIDO2 providers in the wild use non-resident key handles and are thus susceptible to accounts with them being linked to other services by malicious actors. 

In the course of our research we were not able to investigate all available authenticators, however, based on the fact that the vulnerable authenticators we found are manufactured by major security vendors, we expect this flaw to be present in other products as well. What is worse, identifying the vulnerability might not be enough since most hardware tokens do not support firmware updates. Fortunately, we proposed a mitigation mechanism that involves the client side (i.e. browsers) and is independent of the authenticator which prevents remote attacks via web pages. This approach allows the user to update their FIDO client (typically a browser) and still use vulnerable tokens safely for web based authentication via the browser. 

We notified and worked with the affected vendors to secure the privacy of FIDO2 protocol.


\bibliographystyle{./bibliography/IEEEtran}
\bibliography{./bibliography/IEEEexample}

\begin{thebibliography}{10}
\providecommand{\url}[1]{#1}
\csname url@samestyle\endcsname
\providecommand{\newblock}{\relax}
\providecommand{\bibinfo}[2]{#2}
\providecommand{\BIBentrySTDinterwordspacing}{\spaceskip=0pt\relax}
\providecommand{\BIBentryALTinterwordstretchfactor}{4}
\providecommand{\BIBentryALTinterwordspacing}{\spaceskip=\fontdimen2\font plus
\BIBentryALTinterwordstretchfactor\fontdimen3\font minus
  \fontdimen4\font\relax}
\providecommand{\BIBforeignlanguage}[2]{{%
\expandafter\ifx\csname l@#1\endcsname\relax
\typeout{** WARNING: IEEEtran.bst: No hyphenation pattern has been}%
\typeout{** loaded for the language `#1'. Using the pattern for}%
\typeout{** the default language instead.}%
\else
\language=\csname l@#1\endcsname
\fi
#2}}
\providecommand{\BIBdecl}{\relax}
\BIBdecl

\bibitem{DBLP:conf/ccs/ThomasLZBRIMCEM17}
K.~Thomas \emph{et~al.}, ``Data breaches, phishing, or malware?: Understanding
  the risks of stolen credentials,'' in \emph{Proceedings of the 2017 {ACM}
  {SIGSAC} Conference on Computer and Communications Security, {CCS} 2017},
  2017, pp. 1421--1434.

\bibitem{DBLP:conf/soups/KarunakaranTBC18}
S.~Karunakaran \emph{et~al.}, ``Data breaches: User comprehension,
  expectations, and concerns with handling exposed data,'' in \emph{Fourteenth
  Symposium on Usable Privacy and Security, {SOUPS} 2018, Baltimore, MD, USA,
  August 12-14, 2018}, 2018, pp. 217--234.

\bibitem{FIDO2}
\BIBentryALTinterwordspacing
C.~Brand \emph{et~al.}, ``Client to authenticator protocol ({CTAP}),'' 2019.
  [Online]. Available:
  \url{https://fidoalliance.org/specs/fido-v2.0-ps-20190130/fido-client-to-authenticator-protocol-v2.0-ps-20190130.pdf}
\BIBentrySTDinterwordspacing

\bibitem{UAF}
\BIBentryALTinterwordspacing
R.~Lindemann and E.~Tiffany, ``{FIDO UAF} protocol specification,'' 2020.
  [Online]. Available:
  \url{https://fidoalliance.org/specs/fido-uaf-v1.2-ps-20201020/fido-uaf-protocol-v1.2-ps-20201020.pdf}
\BIBentrySTDinterwordspacing

\bibitem{U2F}
\BIBentryALTinterwordspacing
S.~Srinivas, D.~Balfanz, E.~Tiffany, and A.~Czeskis, ``Universal 2nd factor
  ({U2F}) overview,'' 2017. [Online]. Available:
  \url{https://fidoalliance.org/specs/fido-u2f-v1.2-ps-20170411/fido-u2f-overview-v1.2-ps-20170411.pdf}
\BIBentrySTDinterwordspacing

\bibitem{FIDO_ctap_spec}
\BIBentryALTinterwordspacing
(2019, Jan.) Client to authenticator protocol (ctap). [Online]. Available:
  \url{https://fidoalliance.org/specs/fido-v2.0-ps-20190130/fido-client-to-authenticator-protocol-v2.0-ps-20190130.html}
\BIBentrySTDinterwordspacing

\bibitem{Kumar:21:WAA}
\BIBentryALTinterwordspacing
A.~Kumar \emph{et~al.}, ``Web authentication: An {API} for accessing public key
  credentials - level 2,'' W3C, {W3C} Recommendation, Apr. 2021. [Online].
  Available: \url{https://www.w3.org/TR/2021/REC-webauthn-2-20210408/}
\BIBentrySTDinterwordspacing

\bibitem{FIDO_android_announcement}
\BIBentryALTinterwordspacing
(2019, Feb.) Android now fido2 certified, accelerating global migration beyond
  passwords. [Online]. Available:
  \url{https://fidoalliance.org/android-now-fido2-certified-accelerating-global-migration-beyond-passwords/}
\BIBentrySTDinterwordspacing

\bibitem{FIDO_windows_announcement}
\BIBentryALTinterwordspacing
(2019, May) Microsoft achieves fido2 certification for windows hello. [Online].
  Available:
  \url{https://fidoalliance.org/microsoft-achieves-fido2-certification-for-windows-hello/}
\BIBentrySTDinterwordspacing

\bibitem{FIDO_ios_announcement}
\BIBentryALTinterwordspacing
(2020, Jul.) Expanded support for fido authentication in ios and macos.
  [Online]. Available:
  \url{https://fidoalliance.org/expanded-support-for-fido-authentication-in-ios-and-macos/}
\BIBentrySTDinterwordspacing

\bibitem{FIDO_deployment_us_gov}
\BIBentryALTinterwordspacing
(2019, Mar.) U.s. general services administration’s rollout of fido2 on
  login.gov. [Online]. Available:
  \url{https://fidoalliance.org/u-s-general-services-administrations-rollout-of-fido2-on-login-gov/}
\BIBentrySTDinterwordspacing

\bibitem{FIDO_deployment_visa}
\BIBentryALTinterwordspacing
(2019, Jan.) Visa case study. [Online]. Available:
  \url{https://fidoalliance.org/visa-case-study/}
\BIBentrySTDinterwordspacing

\bibitem{FIDO_deployment_nhs}
\BIBentryALTinterwordspacing
(2021, Feb.) National health service uses fido authentication for enhanced
  login. [Online]. Available:
  \url{https://fidoalliance.org/national-health-service-uses-fido-authentication-for-enhanced-login/}
\BIBentrySTDinterwordspacing

\bibitem{FIDO_webauthn_spec}
\BIBentryALTinterwordspacing
(2021, Feb.) Web authentication: An api for accessing public key credentials.
  [Online]. Available: \url{https://www.w3.org/TR/webauthn-2/}
\BIBentrySTDinterwordspacing

\bibitem{4cefd0d46505406a96e10ae7d7afcaae}
C.~Castelluccia \emph{et~al.}, ``\BIBforeignlanguage{English}{Betrayed by your
  ads! reconstructing user profiles from targeted ads},'' in
  \emph{\BIBforeignlanguage{English}{Privacy Enhancing Technologies}}, ser.
  Lecture Notes in Computer Science, S.~Fischer-Hubner and M.~Wright,
  Eds.\hskip 1em plus 0.5em minus 0.4em\relax Springer, 2012.

\bibitem{DBLP:conf/www/GogaLPFST13}
O.~Goga \emph{et~al.}, ``Exploiting innocuous activity for correlating users
  across sites,'' in \emph{22nd International World Wide Web Conference, {WWW}
  '13}, 2013, pp. 447--458.

\bibitem{DBLP:conf/sp/NarayananPGBSSS12}
A.~Narayanan \emph{et~al.}, ``On the feasibility of internet-scale author
  identification,'' in \emph{{IEEE} Symposium on Security and Privacy, {SP}
  2012, 21-23 May 2012, San Francisco, California, {USA}}.\hskip 1em plus 0.5em
  minus 0.4em\relax {IEEE} Computer Society, 2012, pp. 300--314.

\bibitem{DBLP:conf/www/RiedererKCKL16}
C.~J. Riederer \emph{et~al.}, ``Linking users across domains with location
  data: Theory and validation,'' in \emph{Proceedings of the 25th International
  Conference on World Wide Web, {WWW} 2016}, 2016, pp. 707--719.

\bibitem{noauthor_yubico_2021}
\BIBentryALTinterwordspacing
``\BIBforeignlanguage{en-US}{Yubico announcement},'' Mar. 2021. [Online].
  Available:
  \url{https://www.yubico.com/blog/yubico-donates-25000-yubikeys-to-microsoft-accountguard-customers-in-31-countries/}
\BIBentrySTDinterwordspacing

\bibitem{noauthor_google_2021}
\BIBentryALTinterwordspacing
``\BIBforeignlanguage{en-us}{Google announcement},'' Oct. 2021. [Online].
  Available:
  \url{https://blog.google/technology/safety-security/delivering-10000-security-keys-high-risk-users/}
\BIBentrySTDinterwordspacing

\bibitem{fido-security}
{FIDO Alliance}, ``{FIDO Security Reference},''
  \url{https://fidoalliance.org/specs/fido-v2.0-id-20180227/fido-security-ref-v2.0-id-20180227.html},
  2018, [Online; accessed 2-May-2021].

\bibitem{9176}
M.~Dworkin, ``\BIBforeignlanguage{en}{Recommendation for block cipher modes of
  operation: Methods for key wrapping},'' 2012-12-13 2012.

\bibitem{journals/ijisec/JohnsonMV01}
D.~Johnson \emph{et~al.}, ``The elliptic curve digital signature algorithm
  (ecdsa).'' \emph{Int. J. Inf. Sec.}, vol.~1, no.~1, 2001.

\bibitem{abs-1904-10629}
B.~Z.~H. Zhao \emph{et~al.}, ``A decade of mal-activity reporting: {A}
  retrospective analysis of internet malicious activity blacklists,''
  \emph{CoRR}, vol. abs/1904.10629, 2019.

\bibitem{FIDO_privacy_principles}
\BIBentryALTinterwordspacing
F.~Alliance. (2014, Feb.) Privacy principles whitepaper. [Online]. Available:
  \url{https://media.fidoalliance.org/wp-content/uploads/2014/12/FIDO\_Alliance\_Whitepaper\_Privacy\_Principles.pdf}
\BIBentrySTDinterwordspacing

\bibitem{webauth_priv_considerations}
\BIBentryALTinterwordspacing
{W3C}, ``{Privacy Considerations},'' 2021. [Online]. Available:
  \url{https://www.w3.org/TR/webauthn-2/#sctn-privacy-considerations}
\BIBentrySTDinterwordspacing

\bibitem{FIDO_cert_requirements}
\BIBentryALTinterwordspacing
(2022, Feb.) Fido authenticator security requirements. [Online]. Available:
  \url{https://fidoalliance.org/specs/fido-security-requirements/fido-authenticator-security-requirements-v1.5-fd-20211102.html}
\BIBentrySTDinterwordspacing

\bibitem{chromium-fido}
\BIBentryALTinterwordspacing
Google, ``Chromium: implementation of fido2 client.'' [Online]. Available:
  \url{https://chromium.googlesource.com/chromium/src/+/refs/heads/main/device/fido/get_assertion_task.cc}
\BIBentrySTDinterwordspacing

\bibitem{opensk}
\BIBentryALTinterwordspacing
{Google}, ``Opensk: open-source implementation for fido u2f and fido 2 security
  keys.'' [Online]. Available:
  \url{https://github.com/google/OpenSK/blob/5e682d9e176e936c22fcb963a708ffb0b47a33e6/src/ctap/mod.rs}
\BIBentrySTDinterwordspacing

\bibitem{OWASP_top10}
\BIBentryALTinterwordspacing
(2022, Feb.) Owasp top 10. [Online]. Available: \url{https://owasp.org/Top10/}
\BIBentrySTDinterwordspacing

\bibitem{fido_lvl_1}
\BIBentryALTinterwordspacing
{FIDO Alliance}, ``{Authenticator Level 1},'' 2017. [Online]. Available:
  \url{https://fidoalliance.org/certification/authenticator-certification-levels/authenticator-level-1/}
\BIBentrySTDinterwordspacing

\bibitem{8835386}
D.~{Genkin} \emph{et~al.}, ``Synesthesia: Detecting screen content via remote
  acoustic side channels,'' in \emph{2019 IEEE Symposium on Security and
  Privacy (SP)}, 2019, pp. 853--869.

\bibitem{w3c_mediacapture}
\BIBentryALTinterwordspacing
(2021, Mar.) Media capture and streams. [Online]. Available:
  \url{https://www.w3.org/TR/mediacapture-streams/}
\BIBentrySTDinterwordspacing

\bibitem{9152694}
S.~{Ghorbani Lyastani} \emph{et~al.}, ``Is fido2 the kingslayer of user
  authentication? a comparative usability study of fido2 passwordless
  authentication,'' in \emph{2020 IEEE Symposium on Security and Privacy (SP)},
  2020, pp. 268--285.

\bibitem{238317}
S.~Ciolino \emph{et~al.}, ``Of two minds about two-factor: Understanding
  everyday {FIDO} u2f usability through device comparison and experience
  sampling,'' in \emph{15th Symposium on Usable Privacy and Security ({SOUPS}
  2019)}.\hskip 1em plus 0.5em minus 0.4em\relax {USENIX} Association, Aug.
  2019.

\bibitem{cryptoeprint:2020:756}
M.~Barbosa \emph{et~al.}, ``Provable security analysis of fido2,'' in
  \emph{Advances in Cryptology -- CRYPTO 2021}.\hskip 1em plus 0.5em minus
  0.4em\relax Cham: Springer International Publishing, 2021, pp. 125--156.

\bibitem{10.1145/3190619.3190640}
I.~B. Guirat and H.~Halpin, ``Formal verification of the w3c web authentication
  protocol,'' in \emph{Proceedings of the 5th Annual Symposium and Bootcamp on
  Hot Topics in the Science of Security}, ser. HoTSoS '18, 2018.

\bibitem{UAFAnalysis2021}
H.~Feng, H.~Li, X.~P. Pan, and Z.~Zhao, ``A formal analysis of the fido uaf
  protocol,'' in \emph{Proceedings of the Network and Distributed System
  Security Symposium (NDSS)}, 2021.

\bibitem{titan_ninja}
\BIBentryALTinterwordspacing
(2021, Jan.) A side journey to titan side-channel attack on the google titan
  security key. [Online]. Available:
  \url{https://ninjalab.io/wp-content/uploads/2021/01/a_side_journey_to_titan.pdf}
\BIBentrySTDinterwordspacing

\bibitem{10.1145/3319535.3363283}
A.~Alam \emph{et~al.}, ``Poster: Let history not repeat itself (this time) --
  tackling webauthn developer issues early on,'' in \emph{Proceedings of the
  2019 ACM SIGSAC Conference on Computer and Communications Security}, ser. CCS
  '19, 2019.

\bibitem{9343231}
E.~Klieme \emph{et~al.}, ``Fidonuous: A fido2/webauthn extension to support
  continuous web authentication,'' in \emph{2020 IEEE 19th International
  Conference on Trust, Security and Privacy in Computing and Communications
  (TrustCom)}, 2020.

\bibitem{10.1145/3319535.3363258}
D.~Chakraborty and S.~Bugiel, ``Simfido: Fido2 user authentication with
  simtpm,'' in \emph{Proceedings of the 2019 ACM SIGSAC Conference on Computer
  and Communications Security}, ser. CCS '19.\hskip 1em plus 0.5em minus
  0.4em\relax Association for Computing Machinery, 2019, p. 2569–2571.

\bibitem{8835225}
E.~Dauterman \emph{et~al.}, ``True2f: Backdoor-resistant authentication
  tokens,'' in \emph{2019 IEEE Symposium on Security and Privacy (SP)}, 2019,
  pp. 398--416.

\bibitem{10.1145/3372297.3417292}
N.~Frymann \emph{et~al.}, ``Asynchronous remote key generation: An analysis of
  yubico's proposal for w3c webauthn,'' in \emph{Proceedings of the 2020 ACM
  SIGSAC Conference on Computer and Communications Security}, ser. CCS
  '20.\hskip 1em plus 0.5em minus 0.4em\relax Association for Computing
  Machinery, 2020, p. 939–954.

\bibitem{272198}
K.~Pfeffer \emph{et~al.}, ``On the usability of authenticity checks for
  hardware security tokens,'' in \emph{30th {USENIX} Security Symposium
  ({USENIX} Security 21)}.\hskip 1em plus 0.5em minus 0.4em\relax {USENIX}
  Association, Aug. 2021.

\bibitem{255646}
F.~M. Farke \emph{et~al.}, ``{\textquotedblleft}you still use the password
  after all{\textquotedblright} {\textendash} exploring fido2 security keys in
  a small company,'' in \emph{Sixteenth Symposium on Usable Privacy and
  Security ({SOUPS} 2020)}.\hskip 1em plus 0.5em minus 0.4em\relax {USENIX}
  Association, Aug. 2020, pp. 19--35.

\bibitem{solokeys}
\BIBentryALTinterwordspacing
SoloKeys, ``Solokeys: open-source firmware implementation.'' [Online].
  Available: \url{https://github.com/solokeys/solo/blob/master/fido2/crypto.c}
\BIBentrySTDinterwordspacing

\end{thebibliography}

\appendix



\renewcommand{\textfraction}{0.05} 
\clearpage
\section{Responsible Disclosure}
\label{Apdix:vendors}

The results of our research were sent to the responsible parties. We notified three vendors of web browsers (Chromium, Firefox, and Safari), two hardware tokens vendors (Feitian and Hypersecu) and the FIDO Alliance.

For Chromium, we sent the vulnerability report though the chromium bug tracing platform. After a brief explanation,
the chromium security team acknowledged the vulnerability. On 14th September 2021, we received a plan how the mitigation will be implemented: 

\noindent
\textit{"I think the mitigation step we should take is to deduplicate allowedCredentials before making CTAP requests. This should eliminate the amplification, and then the time it takes the user to touch the key would hopefully dominate any difference in CTAP request time. We can also limit the size of that list explicitly, but we would need to be careful not to break any weird outlier sites with users that have lots of enrolled credentials. As for severity, per https://www.chromium.org/developers/severity-guidelines I would say the cross-origin user correlation is a type of information disclosure." - martinkr@google.com} 

\noindent
On 5th October 2021 the first implementation which included a deduplication mechanism combined with a limitation of the size of the allowCredential list (limit set to 64 key handles) was provided. We tested the vulnerability fix on the canary release of Chrome browser and confirmed that the attack is mitigated. The vulnerability received id CVE-2021-38022.

Similarly, Firefox was notified through their bug tracing channel. The Firefox team recognized the issue and assigned internal bug id 1730434. On 15th September 2021, the issue status was changed from unconfirmed to new. At the time of writing, the implementation of the mitigation mechanism is in progress.

In case of the Safari browser, we sent an email to the product security team. On 28th September 2021, we received a confirmation saying: 

\noindent
\textit{"We don't automatically provide status updates on issues as we work on them. We will reach out if we have any questions or need additional details."}. 

\noindent
Our report received tracking number 781222840 and is undergoing an investigation. 

We reached out to Feitian through their official contact email. On 14th September, our report was passed to Feitian's internal team: 

\noindent
\textit{"I forward your report to R\&D team. Our developers are investigating it, we will update you when we came to conclusion." - lena@ftsafe.com } 

\noindent
We explained the issue to the Feitian's engineering team and provided the detailed recording of the testing procedure. The issue is being investigated. 

The report to Hypersecu was sent to their official contact email. We received their acknowledgment on 13th September 2021. We provided an exhaustive description of the problem together with results and recording of our tests. On 15th October 2021 we received a request for additional parameters describing the vulnerable token:

\noindent
\textit{"Could you please use the following tool to send the HyperFIDO info to us? We want to make sure the version info of the key you tested.
On the other hand, would you mind letting us know from where you bought the key? Was it Amazon AU?" - james@hypersecu.com}

\noindent
We provided the requested data. The Hypersecu team is investigating the problem. 

Additionally, we sent our report to the FIDO Alliance Security Certification Secretariat. The Fido Alliance representative analysed our report and endorsed our findings:

\noindent
\textit{"After a quick analysis, I'd like to congratulate you for your successful timing attacks conducted on multiple certified silent authenticators at L1." - roland@fidoalliance.org}

\noindent
Moreover, we learned that our contribution will help to spread awareness of the advantages of the highest certification levels (L3/L3+), which require deep laboratory analysis including timing measurements:

\noindent
\textit{"Indeed, L1 certification is not designed to provide assurance against such attacks so, there isn't much here in terms of actions we could do in addition to notifying the vendors. However, I think that RPs should be made aware of such type of attack so they can better understand why L3/L3+ certification makes sense and probably help in creating incentives in this sense. And that's something where FIDO could potentially help based on your findings." - roland@fidoalliance.org}

\clearpage
\onecolumn
\section{Silent Authentication Measurements}
\label{Apdix:silent-auth-times}
\vspace{-1cm}

\begin{figure}[h!] 
  \begin{subfigure}[b]{0.5\linewidth}
    \centering
    \includegraphics[width=0.75\linewidth]{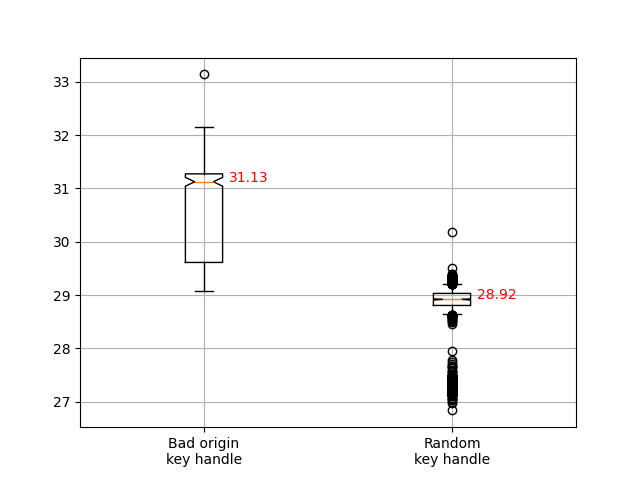} 
    \caption{Silent authentication times for Feitian K26} 
    \label{fig7:a} 
  \end{subfigure}
  \begin{subfigure}[b]{0.5\linewidth}
    \centering
    \includegraphics[width=0.75\linewidth]{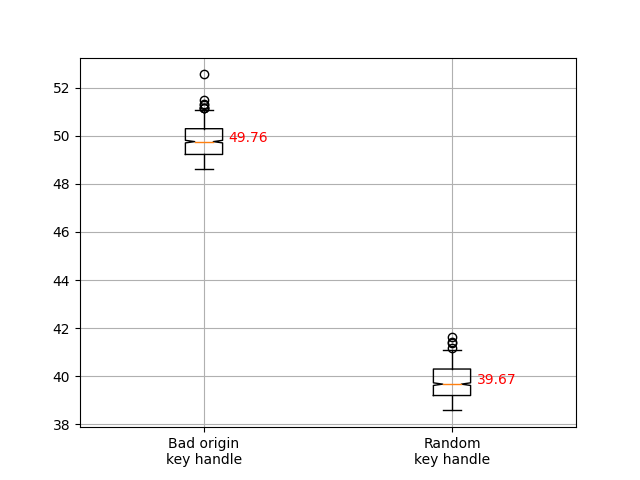} 
    \caption{Silent authentication times for HyperFIDO Titan Pro} 
    \label{fig7:b} 
  \end{subfigure} 
  \begin{subfigure}[b]{0.5\linewidth}
    \centering
    \includegraphics[width=0.75\linewidth]{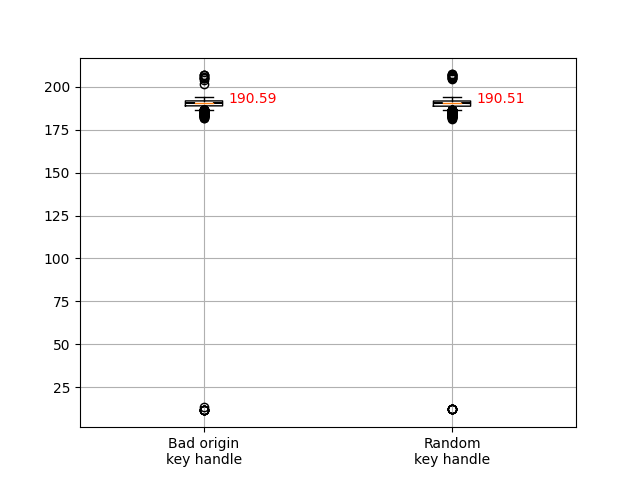} 
    \caption{Silent authentication times for Yubikey 5 \\(samples below 20ms represent initial calls \\ without triggering the defence mechanism)} 
    \label{fig7:c} 
  \end{subfigure}
  \begin{subfigure}[b]{0.5\linewidth}
    \centering
    \includegraphics[width=0.75\linewidth]{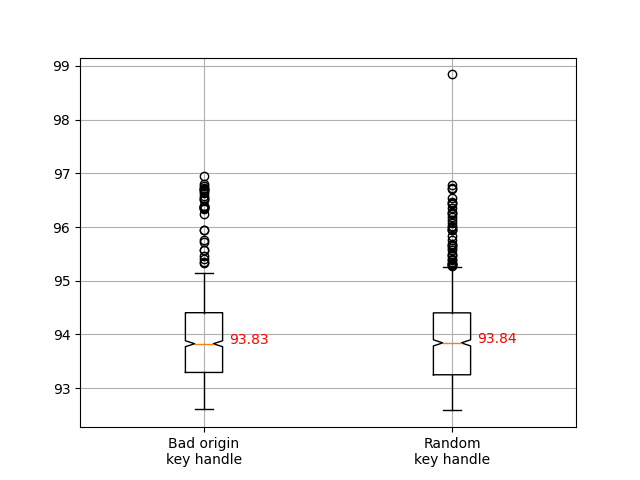} 
    \caption{Silent authentication times for Token2 T2F2 Bio} 
    \label{fig7:d} 
  \end{subfigure} 
    \begin{subfigure}[b]{0.5\linewidth}
    \centering
    \includegraphics[width=0.75\linewidth]{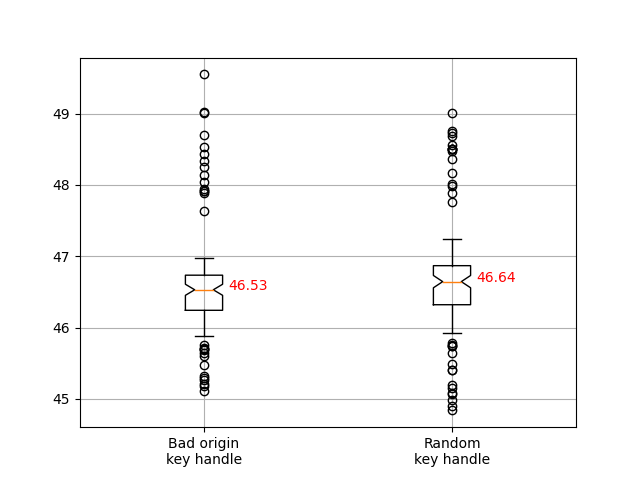} 
    \caption{Silent authentication times for TrustKey G320H} 
    \label{fig7:c} 
  \end{subfigure}
  \begin{subfigure}[b]{0.5\linewidth}
    \centering
    \includegraphics[width=0.75\linewidth]{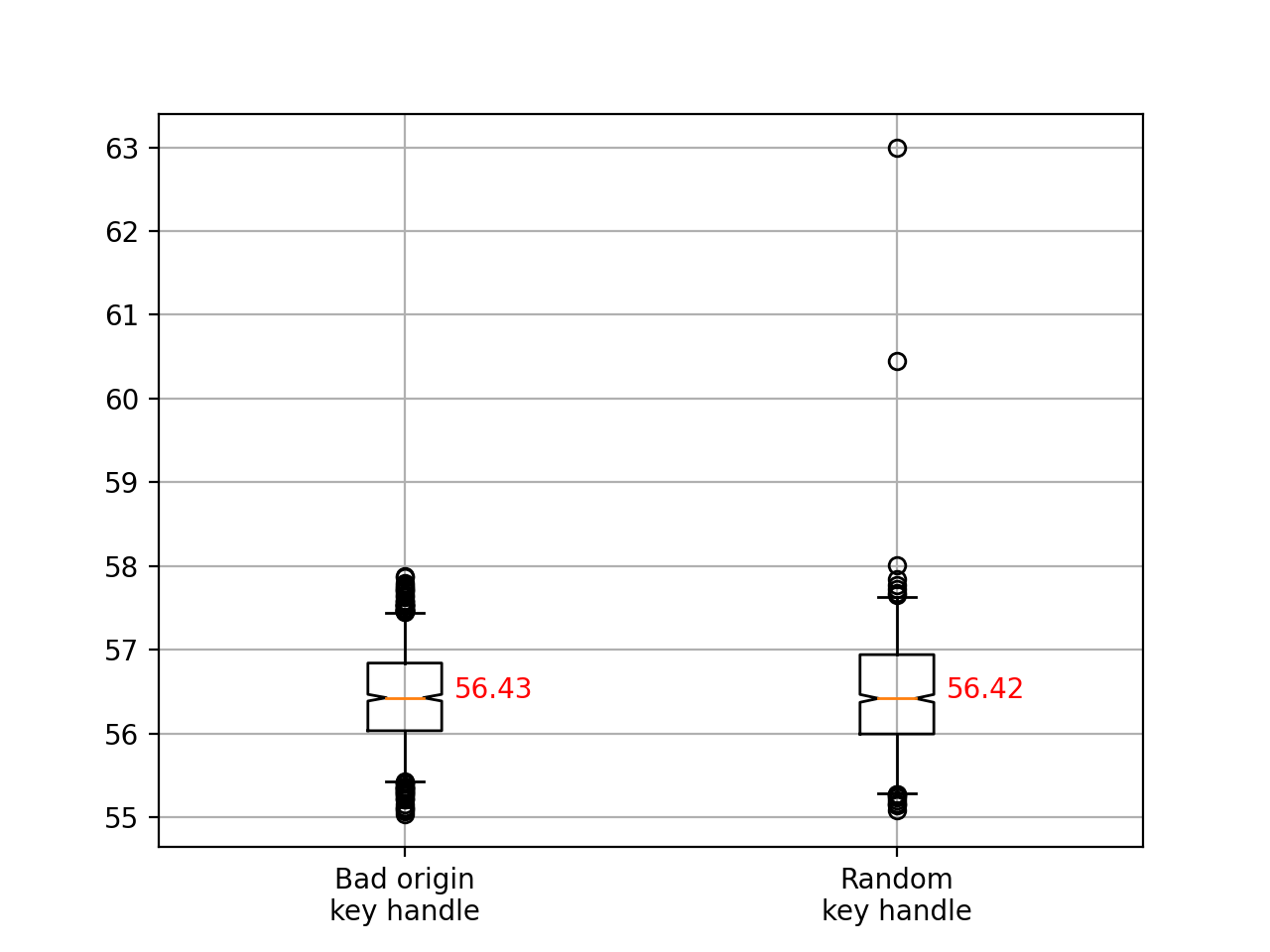} 
    \caption{Silent authentication times for Google Titan} 
    \label{fig7:d} 
  \end{subfigure}
    \begin{subfigure}[b]{0.5\linewidth}
    \centering
    \includegraphics[width=0.75\linewidth]{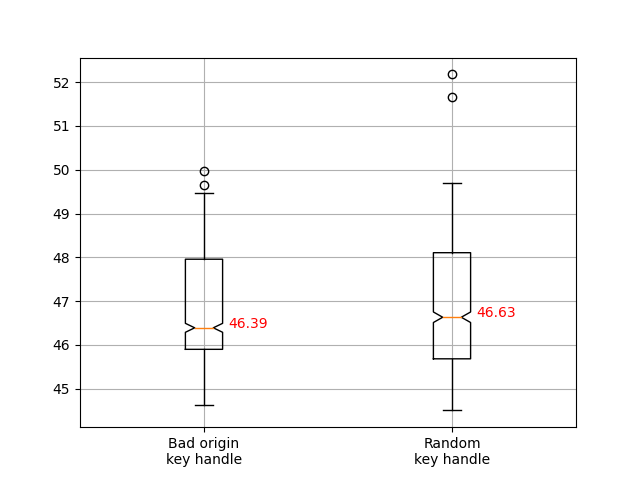} 
    \caption{Silent authentication times for VeriMark Guard Fingerprint} 
    \label{fig7:c} 
  \end{subfigure}
  \begin{subfigure}[b]{0.5\linewidth}
    \centering
    \includegraphics[width=0.75\linewidth]{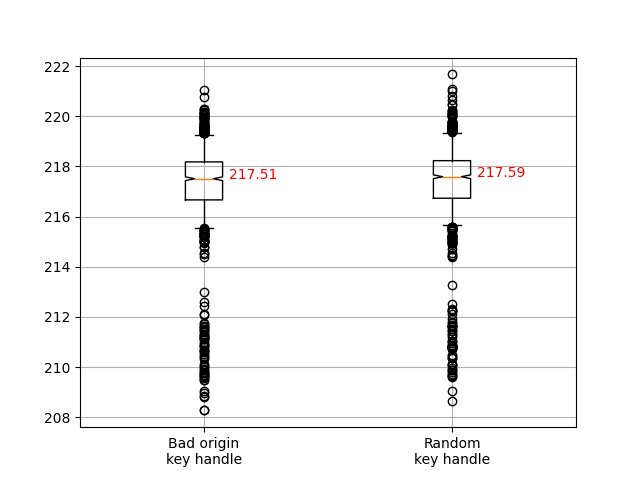} 
    \caption{Silent authentication times for AuthenTrend ATKey.Pro} 
    \label{fig7:d} 
  \end{subfigure}
  \caption{Silent authentication time[ms] measurements}
  \label{fig7} 
\end{figure}
\vspace{-12cm}








\clearpage
\onecolumn
\section{Attack scenarios}\label{appx:attack_scenarios}
\begin{figure}[h!]

    \centering
    \includegraphics[width=\textwidth]{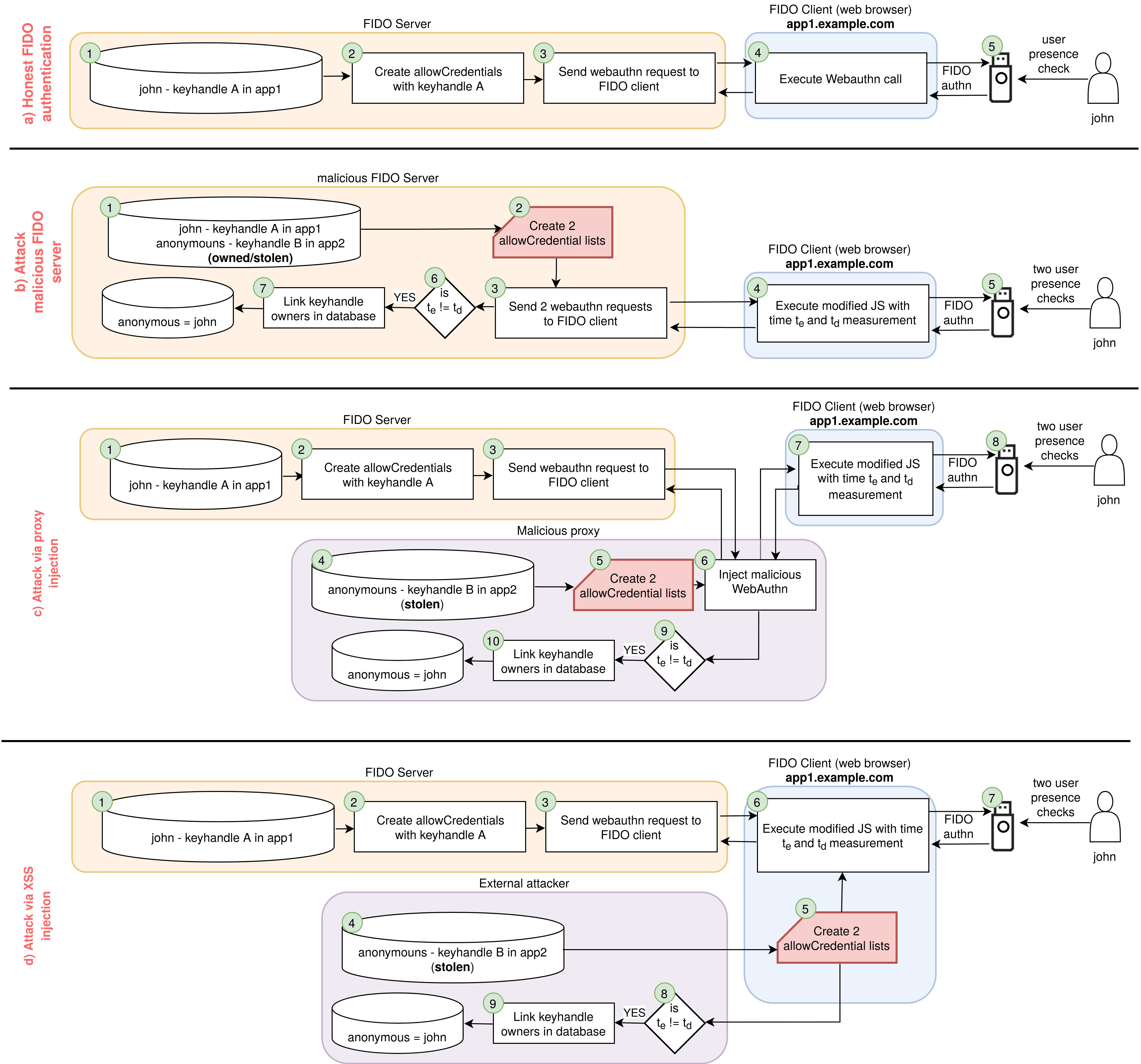}
    \caption{\fix{FIDO timing attack scenarios\\
    \vspace{5pt}
    a) An honest FIDO2 authentication. The FIDO Server uses John's keyhandle (A) to trigger an WebAuthn call in the FIDO Client. \\
    \vspace{5pt}
    b) A malicious FIDO Server, which is in possession of John's keyhandle as well as a keyhandle of an anonymous user. The FIDO server executes the attack to learn if the anonymous user is in fact John. In step 2, two \allCred lists are generated and sent to the FIDO client to execute and measure two consecutive WebAuthn calls (step 4). Then the FIDO server decides if times differ (step 6) and if they do, the identity is linked (step 7).\\
    \vspace{5pt}
    c) An attack by a malicious proxy. In this configuration, the FIDO Server is honest, however the proxy manipulates messages and JavaScript calls to learn the timing difference (step 6). The Javascript execution and linking decision is the same as scenario (b).\\
    \vspace{5pt}
    d) Attack via injection of malicious JavaScript. In this case, an adversary manipulates WebAuthn calls directly from JavaScript (steps 5 and 6). The decision and linking process remain unchanged.
    }}\label{fig:scenarios}
\end{figure}

\clearpage
\twocolumn

\onecolumn
\section{Open source implementations of key handling} \label{apdix:implementation}
\begin{figure}[H]
\begin{lstlisting}[language=Rust,firstnumber=259]
    // Decrypts a credential ID and writes the private key into a PublicKeyCredentialSource.
    // None is returned if the HMAC test fails or the relying party does not match the
    // decrypted relying party ID hash.
    pub fn decrypt_credential_source(
        &self,
        credential_id: Vec<u8>,
        rp_id_hash: &[u8],
    ) -> Result<Option<PublicKeyCredentialSource>, Ctap2StatusCode> {
        if credential_id.len() != CREDENTIAL_ID_SIZE {
            return Ok(None);
        }
        let master_keys = self.persistent_store.master_keys()?;
        let payload_size = credential_id.len() - 32;
        if !verify_hmac_256::<Sha256>(
            &master_keys.hmac,
            &credential_id[..payload_size],
            array_ref![credential_id, payload_size, 32],
        ) {
            return Ok(None);
        }
        let aes_enc_key = crypto::aes256::EncryptionKey::new(&master_keys.encryption);
        let aes_dec_key = crypto::aes256::DecryptionKey::new(&aes_enc_key);
        let mut iv = [0; 16];
        iv.copy_from_slice(&credential_id[..16]);
        let mut blocks = [[0u8; 16]; 4];
        for i in 0..4 {
            blocks[i].copy_from_slice(&credential_id[16 * (i + 1)..16 * (i + 2)]);
        }
        cbc_decrypt(&aes_dec_key, iv, &mut blocks);
        let mut decrypted_sk = [0; 32];
        let mut decrypted_rp_id_hash = [0; 32];
        decrypted_sk[..16].clone_from_slice(&blocks[0]);
        decrypted_sk[16..].clone_from_slice(&blocks[1]);
        decrypted_rp_id_hash[..16].clone_from_slice(&blocks[2]);
        decrypted_rp_id_hash[16..].clone_from_slice(&blocks[3]);
        if rp_id_hash != decrypted_rp_id_hash {
            return Ok(None);
        }
\end{lstlisting}
\caption{Key Decryption Function in OpenSK FIDO Token Implementation \cite{opensk}} \label{fig:opensk}
\end{figure}

\begin{figure}[H]
\begin{lstlisting}[language=C,firstnumber=264]
void generate_private_key(uint8_t * data, int len, uint8_t * data2, int len2, uint8_t * privkey)
{
    crypto_sha256_hmac_init(CRYPTO_MASTER_KEY, 0, privkey);
    crypto_sha256_update(data, len);
    crypto_sha256_update(data2, len2);
    crypto_sha256_update(master_secret, 32);    // TODO AES
    crypto_sha256_hmac_final(CRYPTO_MASTER_KEY, 0, privkey);

    crypto_aes256_init(master_secret + 32, NULL);
    crypto_aes256_encrypt(privkey, 32);
}
\end{lstlisting}
\center
\caption{Pseudorandom Key Generation in SoloKeys Firmware Implementation \cite{solokeys}} \label{fig:solokeys}
\end{figure}


\end{document}